\documentclass[%
 aip,
 amsmath,amssymb,
 reprint, 
twocolumn 
]{revtex4-1}

\usepackage{graphicx}
\usepackage{dcolumn}
\usepackage{bm}
\usepackage{multirow}
\usepackage{bbm}
\usepackage[utf8]{inputenc}
\usepackage[T1]{fontenc}
\usepackage{mathptmx}
\usepackage{etoolbox}
\usepackage{subcaption}
\usepackage{graphicx}
\usepackage[dvipsnames]{xcolor}

\newcommand{\nycm}[1]{}

\newcommand{\nycmnd}[1]{}

\makeatother
\begin{document}

\preprint{AIP/123-QED}

\title[Jinzhen Zhu]{Towards a Universal Foundation Model for Protein Dynamics: A Multi-Chain Tree-Structured Framework with Transformer Propagators}
\author{Jinzhen Zhu}

\email{zhujinzhenlmu@gmail.com}
\affiliation{ 
Shanghai AI Laboratory, Shanghai, 200030, China
}
\date{\today}
\begin{abstract}
Simulating large-scale protein dynamics using traditional all-atom molecular dynamics (MD) remains computationally prohibitive. We present a unified, universal framework for coarse-grained molecular dynamics (CG-MD) that achieves high-fidelity structural reconstruction and generalizes across diverse protein systems. Central to our approach is a hierarchical, tree-structured protein representation (TSCG) that maps Cartesian coordinates into a minimal set of interpretable collective variables. We extend this representation to accommodate multi-chain assemblies, demonstrating sub-angstrom precision in reconstructing full-atom structures from coarse-grained nodes. 
To model temporal evolution, we formulate protein dynamics as stochastic differential equations (SDEs), utilizing a Transformer-based architecture as a universal propagator. By representing collective variables as language-like sequences, our model transcends the limitations of protein-specific networks, generalizing to arbitrary sequence lengths and multi-chain configurations. The framework achieves an acceleration of over 10,000 to 20,000 times compared to traditional MD, generating microsecond-long trajectories within minutes. Our results show that the generated trajectories maintain statistical consistency with all-atom MD in RMSD profiles and structural ensembles. This universal model provides a salable solution for high-throughput protein simulation, offering a significant leap toward a foundation model for molecular dynamics.
\end{abstract}
\maketitle
\section{\label{sec:introduction}Introduction}
Molecular dynamics (MD) simulations serve as a cornerstone for understanding protein structure and function, playing a pivotal role in drug design and protein engineering~\cite{Hollingsworth2018}. However, the computational demand of simulating large biomolecular systems over physiologically relevant timescales remains a significant bottleneck~\cite{Zimmerman2016}. While hardware advancements, such as high-performance Graphics Processing Units (GPUs)~\cite{Bergdorf2021} and specialized supercomputers like Anton~\cite{Shaw2021}, have provided substantial relief, algorithmic improvements—specifically enhanced sampling and coarse-grained (CG) MD—offer a complementary paradigm for acceleration~\cite{Bernardi2015}.

CG methods, which represent groups of atoms as simplified interaction centers, provide a compelling balance between computational efficiency and physical accuracy. These methods typically follow two paradigms: "top-down," driven by experimental thermodynamics, and "bottom-up," which derives effective interactions from high-resolution atomic models~\cite{Noid2013, Liwo2001}. In recent years, machine learning (ML) has revolutionized bottom-up CG simulations by using deep neural networks (DNNs) to approximate complex potential energy surfaces~\cite{Zhang2018, Wang2022}. Despite these strides, maintaining high structural fidelity while achieving a universal, protein-agnostic propagator remains an open challenge.

A critical limitation in many CG representations is the reliance on torsion angles alone. While dihedral fluctuations dominate protein folding, electron orbital hybridization dictates preferred bond-angle geometries (e.g., sp3 or sp2) that exhibit subtle yet structurally vital deviations from ideal values. In traditional torsion-only models, these errors accumulate along the polypeptide chain, leading to unphysical backbone conformations and necessitating the use of Cartesian coordinates for high-precision tasks—as seen in frameworks like AlphaFold 2~\cite{Jumper2021}.

In this work, we introduce a unified framework that overcomes these limitations through a tree-structured protein representation and a Transformer-based propagator. Our approach establishes a bidirectional mapping between Cartesian coordinates and a minimal set of interpretable collective variables (CVs) that account for all heavy atoms in both backbones and side chains. By incorporating both bond and torsion angles into a tree-structured hierarchy, we eliminate cumulative error and allow for near-native reconstruction of complex protein topologies.

Crucially, we extend this representation to handle multi-chain systems, moving beyond the single-chain constraints of earlier models. While our previous iterations utilized protein-specific DNN architectures and Real-NVP generators to approximate stochastic differential equations (SDEs)~\cite{Zhu2024}, we here present a transition toward a more generalizable architecture. By treating protein CVs as "language-like" sequences, we leverage a Transformer framework that is inherently independent of protein size, sequence length, or the number of chains.

We evaluate the performance of this Transformer-based model against the DNN+RealNVP baseline across a diverse set of systems, including the multi-chain protein 3sj9~\cite{Lu2011} and 1bom~\cite{Nagataetal1995}, and the single-chain proteins T1027~\cite{Huangetal2021} and 1l2y~\cite{Neidigh2002}. Our results demonstrate that the Transformer framework not only matches the accuracy of system-specific models but also provides superior generalizability and extrapolation capabilities. By training on a diverse dataset of MD trajectories, this unified propagator achieves a $10^4$-fold speedup, offering a path toward a universal AI model capable of simulating the dynamics of any protein system regardless of its sequence or complexity.
\section{\label{sec:methods}Methods}

\subsection{\label{sec:methods-cv}Collective Variables and Coordinate Transformation}

Our protein representation begins with a general description of the atomic coordinate hierarchy in chemical compounds. We adopt a non-standard notation for clarity: the symbol $\psi$ represents any dihedral angle and $\varphi$ represents any bond angle. 

According to Parsons et al.~\cite{Parsons2005}, the operation for rotating an angle $\theta$ along a normalized rotation vector $\vec{u}=[\vec{u}_x,\vec{u}_y,\vec{u}_z]^T$ is given by the matrix $\hat{R}(\vec{u},\theta)$ (the full expansion of this matrix is provided in Sec.~\ref{sec:coord-transform}). The general formula for coordinate transformation is expressed as a $4\times4$ matrix, where $\vec{T}$ is a length-3 vector representing translation:
\begin{equation}\label{eq:M-gl}
\hat{M}(\vec{u},\theta,\vec{T})=\begin{bmatrix}
\hat{R}(\vec{u},\theta) & \vec{T} \\ 
\vec{0}^T & 1
\end{bmatrix}.
\end{equation}

Converting the local coordinates $\vec{x}_l$ to the global coordinates $\vec{x}_g$ (or converting a local axes frame to a parent axes frame) follows the conversion:
\begin{equation}\label{eq:coord-trans}
    \begin{bmatrix}
        \vec{x}_g\\
        1
    \end{bmatrix}=\hat{M}(\vec{\varphi}, \vec{T})\begin{bmatrix}
        \vec{x}_l\\
        1
    \end{bmatrix}.
\end{equation}

As is illustrated in the upper figure of Fig.~\ref{fig:illustration}, transforming parent axes ($xyz$) data  to local axes ($x'y'z'$) data includes three specific steps, written as the operator $\hat{O} = \hat{M}_3\hat{M}_2\hat{M}_1$: first, $\hat{M}_1$ translates the frame along the x-axis by bond length $l_b$, $\hat{R}(\hat{u},\theta)=\hat{I}$ with $\hat{I}$ being the identity matrix and $\vec{T}=[l_b,0,0]^T$; second, $\hat{M}_2$ rotates the frame along the new z-axis by bond angle $\varphi$ with $\theta=\varphi$, $\vec{T}=\vec{0}$ and $\vec{u}=[0,0,1]^T$; third, $\hat{M}_3$ rotates the frame along the new x-axis by dihedral angle $\psi$ with $\theta=\psi$, $\vec{T}=\vec{0}$ and $\vec{u}=[1,0,0]^T$.

For complex molecules like proteins, multiple coordinate frames are used. We denote the global operator for converting current coordinates to the "global" reference as a recursive multiplication:
\begin{equation}\label{eq:ppmin1}
    \hat{P}_I = \hat{P}_{I-1}\hat{O}_I.
\end{equation}
Consequently, the equation $\vec{X}=\hat{P}_I\vec{X^0}$ can be readily applied to determine any protein atom's Cartesian coordinates within the global reference frame.

\subsection{\label{sec:methods-tree}Tree Data Structure Representation}

\begin{figure}
\centering
\begin{subfigure}{0.48\textwidth}
    \includegraphics[width=\textwidth]{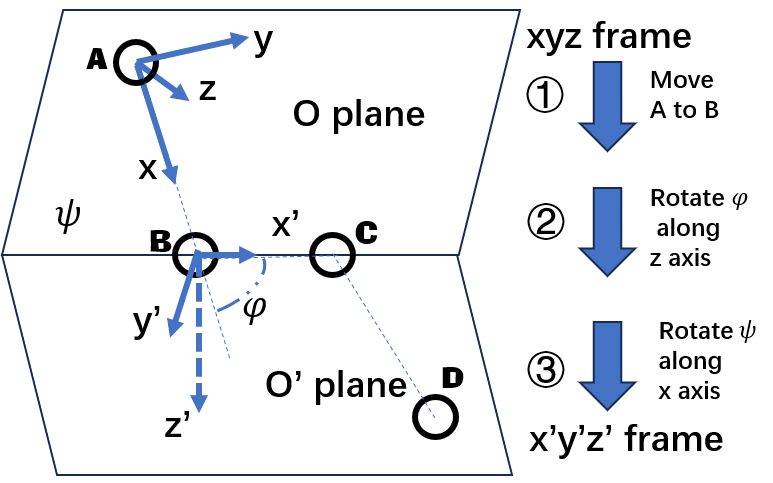}
\end{subfigure}
\begin{subfigure}{0.48\textwidth}
    \includegraphics[width=\textwidth]{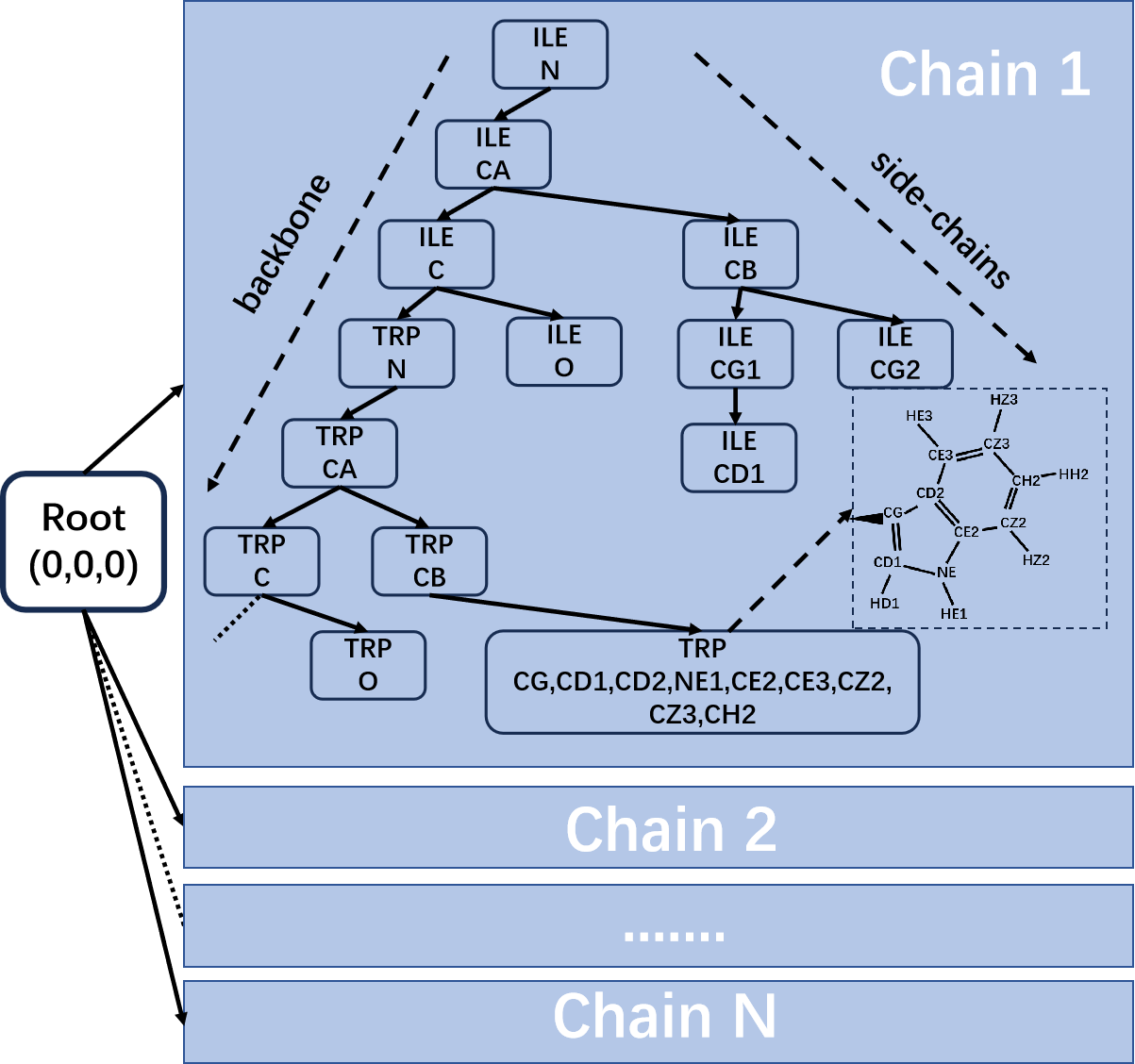}
\end{subfigure}
\caption{This illustration depicts coordinate transform (upper) and tree structure of CVs of atoms in proteins (lower) illustration. Top panel presents four atoms defining two planes with transformations described by $\psi$ and $\varphi$. In the lower panel, an artificial protein with first two amino acids being ILE and TRP is applied to illustrate the tree-structure.}
\label{fig:illustration}
\end{figure}

Recursive data structures naturally lend themselves to representation using tree structures. Eq.~\ref{eq:coord-trans} is universal both for the coordinate transformations between chains and the global protein, and those among different hierarchies in the tree structure of a single chain. 

As illustrated in the lower pannel of Fig.~\ref{fig:illustration}, the data representation among the chains is treated as hierarchies in a tree. The root node is selected as the global origin (0, 0, 0) for convenience, and its children are the roots (the first $N$ atoms) of the respective chains. Within each chain, the tree structure efficiently represents the protein hierarchy and geometry, where each node serves as a local reference frame storing the local coordinates of its constituent atoms. 

This approach builds upon established recursive data structures previously utilized in quantum physics~\cite{Zhu2021,Zhu2020,Zhu2020b}. 
The tree structure efficiently encodes geometric information, where each node inherits a transformation operator ($\hat{P}_I$) from its parent node ($\hat{P}_{I-1}$), as defined in Eq.~\ref{eq:ppmin1}. 
This inheritance mechanism naturally reflects the hierarchical dependencies within the molecular framework. 
Typically, each node represents an individual heavy atom; however, atoms constituting rigid rings (e.g., the $C_\gamma \dots CH_2$ moiety in TRP) are grouped within a single node. 
Throughout the protein dynamics, all bond lengths are treated as constants, as illustrated in the lower panel of Fig.~\ref{fig:illustration} (for the complete structural details of Isoleucine and Tryptophan, see Fig.~\ref{fig:ile-trp}). 
By acknowledging this inherent molecular rigidity, the representation minimizes redundant parameters and significantly enhances data storage efficiency. 
Consequently, a unique hierarchical topology is assigned to each of the 20 standard amino acid types, as summarized in Table~\ref{tab:amino-hierarchy}. 

The recursive matrix multiplication pattern is stored within a fixed tree structure that remains constant throughout the propagation. 
While these recursive operations are computationally demanding, the reconstruction of three-dimensional coordinates is only necessary during trajectory analysis, rather than during the real-time molecular dynamics simulation itself. 
Furthermore, the reconstruction process is highly parallelizable across CPU and GPU architectures, effectively mitigating potential performance bottlenecks.

\subsection{\label{sec:methods-transformer}Transformer-Based Sequence Representation}
\subsubsection{Basic Representation}
For all Cartesian coordinate transformations, Eq.~\ref{eq:ppmin1} can be written as
\begin{equation}
    \mathit{R} = \mathcal{P}(\vec{\theta}, \mathit{R}^0)
\end{equation}
for $N$ Cartesian coordinates
$\mathit{R}=[\vec{x}_1,\vec{x}_2,\dots,\vec{x}_N]^T$
with their corresponding constant local coordinates
$\mathit{R}^0=[\vec{x^0}_1,\vec{x^0}_2,\dots,\vec{x^0}_N]^T$,
$M$ pairs of CVs
$\vec{\theta}=[\psi_1,\psi_2,\cdots,\psi_M,\varphi_1,\varphi_2,\cdots,\varphi_M]^T$,
and the overall operator $\mathcal{P}$ being a combination of all the global operators $\hat{P}_I$.
The inverse transformation for obtaining all angles is written as
\begin{equation}
    \vec{\theta}=\hat{\Theta}(\mathit{R}).
\end{equation}
To overcome the periodicity of angles, in the network we project angles to sine-cosine values as
\begin{equation}\label{eq:S-vec}
    \bold{S}=\mathbb{P}(\vec{\theta})=[\cos\psi_1,\cdots,\cos\psi_M,\sin\varphi_1,\cdots,\sin\varphi_M]^T.
\end{equation}
For the simpliest case, one may represent the CVs at time stamp $t$ with a 1-D vector $S_t$. This saves the memory but the size is dependent on the sequence, whose operator size is also variant, which limits its applications. We will detail this later. 
\subsubsection{Linguistic Sequences Representation}
By representing protein CVs as linguistic sequences, we can effectively integrate Transformer architectures into our learning process. Consider a $C$-chain protein compound, where each chain $c$ has a sequence length $N_c$. The collective variable $S_t$ at time $t$ is represented as a matrix of dimensions $[2+\sum_1^C N_c] \times 2L$, where $L$ is a constant. Physically, $L$ is chosen to encompass the full set of angular variables required for any standard amino acid, such that $2L \geq 40$.

This data matrix is structured as:
\begin{equation}\label{eq:St}
    S_t=\begin{bmatrix}
        \vec{T}^T \\
        \vec{\theta}_f^T\\
        \hat{\theta}_1\\
        \vdots\\
        \hat{\theta}_C
    \end{bmatrix},
\end{equation}
where the first two rows ($\vec{T}^T$ and $\vec{\theta}_f^T$) contain the translational origins and rotational angles of the chains, transformed into a sine-cosine format for periodicity (the detailed mathematical expansions of these frame rows are provided in Sec.~\ref{sec:frame-normalization}). 

The subsequent rows encode the information specific to each protein chain. Within $\hat{\theta}_c$, the $i$-th row stores the dihedral and bond angles of the $i$-th amino acid:
\begin{equation}\label{eq:thetac}
    \hat{\theta}_c=
    \begin{bmatrix}
        \cos\vec{\theta^c_{1}} & \vec{0}_{L-K_1^c} & \sin\vec{\theta^c_{1}} & \vec{0}_{L-K_1^c}\\
        \vdots & \vdots & \vdots & \vdots \\
        \cos\vec{\theta^c_{N_c}} & \vec{0}_{L-K^c_{N_c}} & \sin\vec{\theta^c_{N_c}} & \vec{0}_{L-K^c_{N_c}}
    \end{bmatrix}.
\end{equation}

\subsubsection{\label{sec:methods-encoding}Positional Encoding}
As with most Transformer models, positional encoding is incorporated to capture the position of each amino acid. The positional encoding matrix, $P$, has the same dimensions as $S_c$ and uniquely incorporates both the amino acid index $p(i)$ and the amino acid type index $pt(i)$:
\begin{equation}
P_{i,j}=
\begin{cases}
\cos\frac{p(i)+N_{max}c}{f^{(j-1)(pt(i)/100+1)/L}}& \text{ if } j\leq L \\
\sin\frac{p(i)+N_{max}c}{f^{(j-1-L)(pt(i)/100+1)/L}}& \text{ if } j> L
\end{cases},
\end{equation}
The positional encoding operator $\mathbb{P}$ is applied to the coefficient matrix $S_t$ as follows:
\begin{equation}
\begin{split}
S^{P}_t&=\mathbb{P}S_t=S_t+P\\
S_t&=\mathbb{P}^{-1}S^P_t=S^{P}_t-P
\end{split},
\end{equation}
Note that after this inverse operation, the left and right halves of the angular components in $S_t$ are not necessarily sine and cosine values of the same angles, leading to unphysical behavior. Moreover, the positional encoding does not account for the zero-filled sub-vectors of $S_t$, which consequently become non-zero and also result in unphysical states $S_t$. Therefore, after applying the inverse positional encoding operator, a masking operator and a normalizing operator are required; we will describe these in the framework section later.

\subsection{\label{sec:method-ff}Propagation}
The propagation could be expressed by the stochastical differential equation (SDE), which is typically expressed as
\begin{equation}\label{eq:general-SDE}
\begin{split}
    \frac{dx(\tau)}{d\tau}&=f(x(\tau))+\sum_\alpha g_\alpha(x(\tau))\xi_\alpha
\end{split}
\end{equation}
where $x(\tau)$ represents the position within the system's phase or state space at time $\tau$. Here, $f$ is a flow vector field or drift force that signifies the deterministic evolution law, while $g_\alpha$ is a collection of vector fields that define the system's coupling to Gaussian white noise $\xi_\alpha$\cite{Slavik2013}. After transformations, whose details are shown in Sec.~\ref{sec:SDE-CV-app}, the progagation could be performed by neural networks represented by CVs.
\subsubsection{General SDE propagator}\label{sec:SDE-CV}
The propagation of collective variable $S_{t+i}$, walking $i$ steps from time $t$ to $t+i$, is expressed to resemble a SDE\cite{Slavik2013}:
\begin{equation}\label{eq:sde-network-formula}
S_{t+i}=
\overset{i}{\overbrace{\mathbb{F}_0\circ\mathbb{F}_0\cdots\circ\mathbb{F}_0}}(S_{t})+\mathbb{P}(\epsilon_{t,i}),
\quad \epsilon_{t,i}\equiv\epsilon_{t,i}\left(S(t)\right),
\end{equation}
where $\mathbb{F}_0$ is the deterministic drift force operator for one step, 
$\overset{i}{\overbrace{\mathbb{F}_0\circ\cdots\circ\mathbb{F}_0}}(S_t)$ its $i$-step composition,
and $\mathbb{P}(\epsilon_{t,i})$ is the noise term, with $\epsilon_{t,i}$ denoting the noise amplitude dependent on the CVs at time $t$ and step size $i$.
Details of the derivation can be found in Sec.~\ref{sec:SDE-overview} in the appendix.
\par
We employ a neural network $\mathbb{F}$ to approximate the drift force $\mathbb{F}_0$.
Without the noise term, Eq.~\ref{eq:sde-network-formula} reduces to the drift force propagator.
We have shown in Sec.~\ref{sec:SDE-drift} that when the loss
\begin{equation}\label{eq:loss-general}
    L_{T_n}=\log\frac{1}{n}\sum_{i=1}^{n}\frac{1}{T-i} \sum_{t=1}^{T-i} \left \Vert \mathbf{S}_{t+i}- \overset{i}{\overbrace{\mathbb{F}\circ \cdots\circ \mathbb{F}}}  (\bold{S}_{t})  \right \Vert^2, \quad 1\leq n\leq T-1,
\end{equation}
reaches its minimum, the trained network effectively approximates the drift force $\mathbb{F}_0$.
Note that Eq.~\ref{eq:loss-general} collapses to the single-step logarithmic MSE loss with $n=1$.
Note that here the CVs $S_t$ could be any dimension, as the values are sumed-up to form the entire loss.

\subsubsection{System specific Propagator}
In our previous work~\cite{Zhu2024}, when the CVs are simply represented as a 1-D vector, as is shown in Eq.~\ref{eq:S-vec} the drift force was modeled by a simple DNN:
\begin{equation}\label{eq:all-network}
    \bold{S}_{t+1}= \mathbb{F}(\bold{S}_{t}) =\mathbb{N}\circ \mathbb{L}_{N_h}\circ\cdots\circ \mathbb{L}_{1}(\bold{S}_{t}),
\end{equation}
where $N_h$ denotes the number of hidden layers, $\mathbb{N}$ is a normalization constraint for the sine-cosine operation, and each layer $\mathbb{L}_d$ follows
\begin{equation}\label{eq:linear-transform}
    \mathbf{s}_d=\mathbb{L}_d(\mathbf{s}_{d-1})=\mathbb{A}(\mathbf{W}_d\mathbf{s}_{d-1}+\mathbf{b}_{d-1}),
\end{equation}
with weights $\mathbf{W}_d\in \Re^{M_d\times M_{d-1}}$, biases $\mathbf{b}\in \Re^{M_d}$, and activation functions $\mathbb{A}$.
The extra noise part is generated by a modified RealNVP framework, for details, refer to Sec.~\ref{sec:SDE-drift} and~\ref{sec:realnvp-noise} in the appendix.
While this proved efficient and accurate, the network size depends on the input protein system, making trained models non-transferable to other proteins.
\subsubsection{Universal Propagator with transformer}
\begin{figure}
\includegraphics[width=0.5\textwidth]{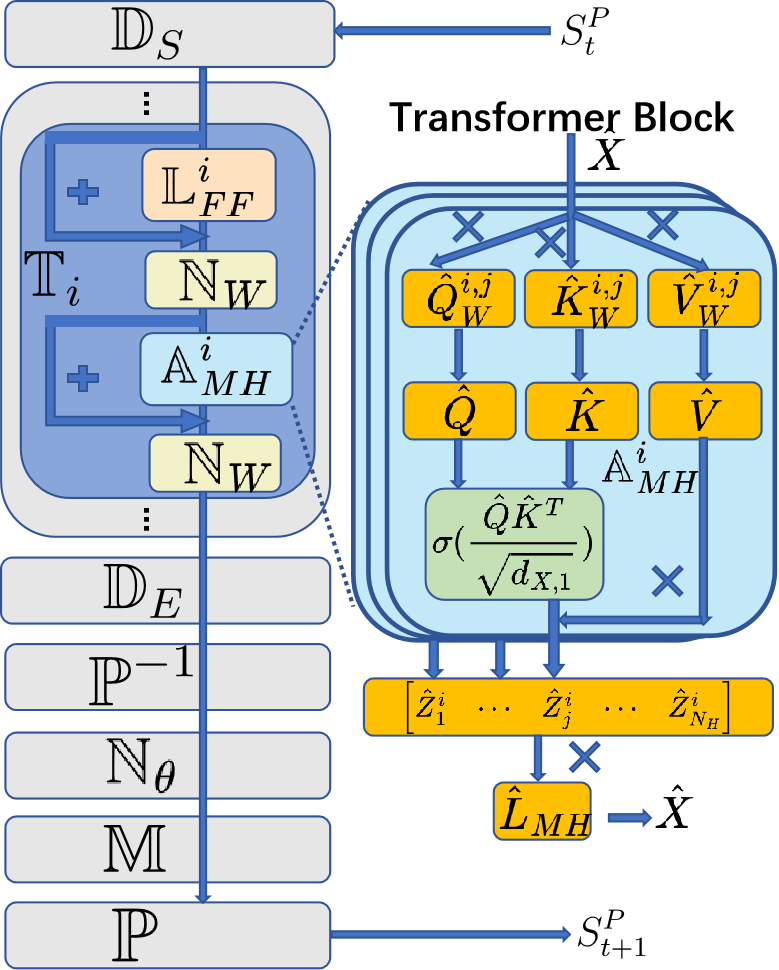}
\caption{Transformer-based propagator architecture. The input CVs $S_t$ are processed by positional encoding $\mathbb{P}$, pre-processing network $\mathbb{D}_S$, a stack of $N_T$ transformer layers, post-processing network $\mathbb{D}_E$, normalization $\mathbb{N}_\theta$, and masking $\mathbb{M}$ to produce $S_{t+1}$.}
\label{fig:transformer}
\end{figure}

The new collective variable representation in Eq.~\ref{eq:St} has a fixed second dimension $L$, making it well-suited for a Transformer architecture~\cite{Vaswani2017}, which has been successfully applied in natural language processing (NLP) and also in MD simulations~\cite{Klein2023}.
By analogy to sequence-to-sequence translation, the input $S_t$ is treated as a sentence of $2+\sum_1^C N_c$ tokens (words), each embedded to a vector of length $2L$, encoding the structural and positional properties of individual amino acids and frame-level properties.
Crucially, unlike standard transformer inputs, no additional embedding operation is needed here: each row of our representation already encodes both positional and type information (via $\hat{S}_c$, Eq.~\ref{eq:St}), and the local structure of each amino acid is fully described by $\hat{\theta}_c$ (Eq.~\ref{eq:thetac}). The encoded state matrix $S^{P}_t$ is used in the operator. If one needs to obtain the generated trajectories in cartesian coordinates, the same substraction, normalization and masking operator can be applied, which process is totally parallel.
\par
The total propagator, as is shown in Fig.~\ref{fig:transformer}, is composed of a sequence of $N_T$ transformer layers $\mathbb{T}_i$, preceded by an inverse positional encoding operator $\mathbb{P}^{-1}$ and followed by a normalization operator $\mathbb{N}_{\theta}$, a masking operator $\mathbb{M}$.
Two fully connected networks, $\mathbb{D}_S$ (pre-processing) and $\mathbb{D}_E$ (post-processing), with fixed input/output dimensions $2L$, are added before and after the transformer to extract relevant features and back-transform the output.
The full propagator is:
\begin{equation}\label{eq:propagator-drift}
    \mathbb{F}_0=\mathbb{P}\circ\mathbb{M}\circ\mathbb{N}_{\theta}\circ\mathbb{P}^{-1}\circ\mathbb{D}_E\circ\mathbb{T}_1\circ\cdots\circ\mathbb{T}_{N_T}\circ \mathbb{D}_S.
\end{equation}
The normalization operator normalizes the $\sin$ and $\cos$ values in each row of $S_t$, including amino acid angles, frame angles, and frame origins (converted to $\sin$/$\cos$ format as shown in Eq.~\ref{eq:frame-prop-origin} and~\ref{eq:frame-prop-angle}).
The mask multiplies a predefined matrix of the same shape as $S_t$, with elements equal to one except at positions occupied by $\vec{0}$ in Eq.~\ref{eq:thetac}, which are set to zero.
\subsubsection{Noise and Stochasticity}\label{sec:noise}

The SDE formulation in Eq.~\ref{eq:sde-network-formula} explicitly includes a noise term $\mathbb{P}(\epsilon_{t,i})$, which is essential for sampling the correct statistical distribution of trajectories.
In our previous single-chain framework~\cite{Zhu2024}, this noise was modeled explicitly using a RealNVP-based noise generator $\mathbb{G}$~\cite{Dinh2017}, with rich physical meaning: the residual noise between successive frames is mapped to a standard normal distribution, and the noise generator is trained by maximizing the Metropolis acceptance ratio.
The full description of this approach, including the network architecture and training objective, is provided in Appendix~\ref{sec:realnvp-noise}.

\par
Within the current Transformer-based unified framework, the direct integration of an explicit noise generator presents significant architectural challenges. 
As an alternative, we introduce stochasticity during inference by leveraging the \emph{dropout} mechanism. 
By randomly deactivating a subset of neuron outputs during each forward pass, dropout serves as a stochastic noise source that allows the model to explore a broader distribution of potential conformational trajectories. 
The dropout rate thus acts as a tunable parameter to calibrate the level of introduced stochasticity. 
Furthermore, maintaining dropout during the training phase ensures that the model effectively learns the stochastic components of the MD trajectory, thereby imbuing the propagation loss with physical relevance. 

While less explicitly derived from first principles than a RealNVP-based noise term, this approach provides a practical and computationally efficient means of incorporating stochastic dynamics into the unified framework. 
When combined with the learned drift force, it enables the robust exploration of diverse trajectory outcomes, as demonstrated in the Results section.

\subsection{Training and Model Assessment}
A key advance over prior work is the shift from protein-specific DNN models to this unified Transformer-based architecture, enabling a single CG model to be trained on and applied to a variety of proteins.
The model is trained on a combined dataset of trajectories from multiple proteins, effectively learning a universal representation of protein dynamics.
\par
An adaptive learning rate strategy is employed during training: the learning rate is dynamically reduced when the loss reaches a plateau.
To reduce computational cost, positional encoding is applied to the entire training dataset prior to training, allowing the $\mathbb{P}$ and $\mathbb{P}^{-1}$ transformations in Eq.~\ref{eq:propagator-drift} to be omitted during the training process.

\section{\label{sec:application-total}Application and Results}
We evaluated our framework across several representative protein systems. 
The fidelity of the tree-structured representation was first assessed through the structural reconstruction of 3sj9~\cite{Lu2011} (a long, multi-chain protein) and T1027~\cite{Huangetal2021} (a long, single-chain protein from the CASP14 dataset~\cite{Kryshtafovychetal2021}). 
Subsequently, to demonstrate the universality of the Transformer-based propagator, we utilized the single-chain protein 1l2y and the two-chain protein 1bom for training; the corresponding reconstruction profiles are also included. 
These smaller systems were selected for initial training to accommodate current hardware constraints, as the memory requirements for modeling 3sj9 exceed the capacity of a single RTX 4060Ti (16GB) GPU. 
We anticipate, however, that scaling to larger proteins and more extensive trajectories will be feasible with more advanced computational resources. 
Molecular dynamics simulations were performed using GROMACS (version 2022.2)\cite{VanDerSpoel2005,Lindahl2001,Berendsen1995} with a 2~fs timestep and a trajectory sampling interval of 0.1~ns.
\subsection{\label{sec:application-structure}Protein structure re-construction}

\begin{figure}
\centering
\begin{subfigure}{0.48\textwidth}
\includegraphics{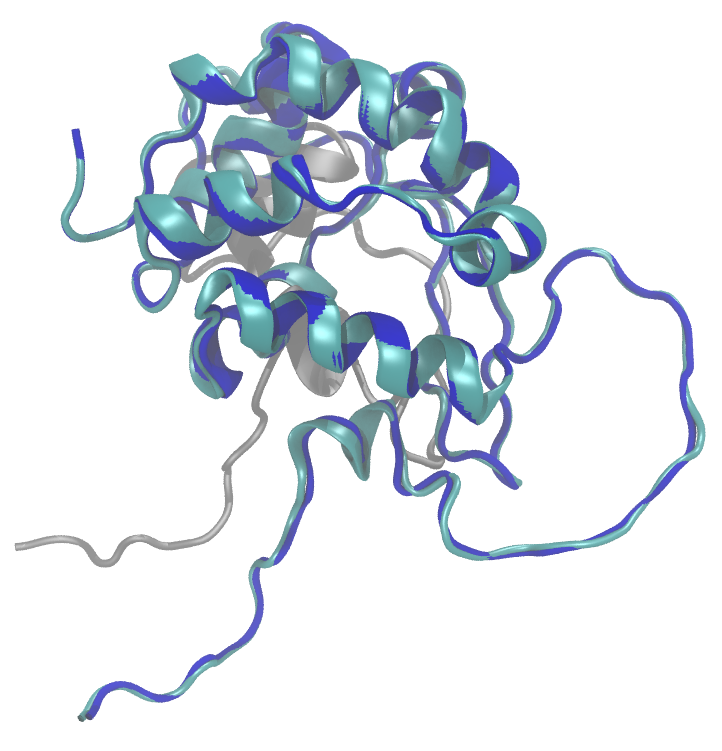}
\end{subfigure}
\begin{subfigure}{0.48\textwidth}  
\includegraphics{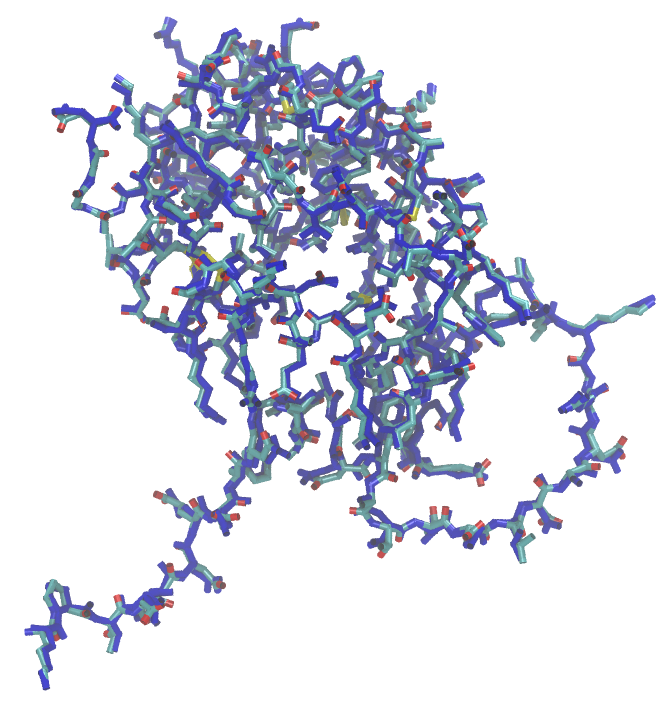}
\end{subfigure}
\centering
\begin{subfigure}{0.2\textwidth}
\includegraphics{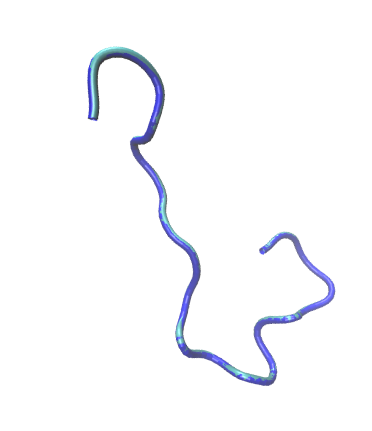}
\end{subfigure}
\caption{Cartesian coordinates reconstruction using dihedral and bond angles. The figures compare the tertiary structures (a and c) and the secondary structures that show side-chains consistency (b). The (a and b) show the comparison for large protein T1027 (168 amino acids) and (c) shows a smaller protein 1l2y (20 amino acids). The original 3D structure is colored blue and the cyan structure is constructed by all real dihedral and bond angles. The gray tertiary structure on the upper figure is constructed by all the real angle data except $sp^3$ bond angles are fixed to 1.29 rads ($109^{\circ}28'$).}
\label{fig:single-chiain-structure}
\end{figure}

For single-chain protein reconstruction, we utilize the 168-residue protein T1027 as a representative case, with results illustrated in the upper panels of Fig.~\ref{fig:single-chiain-structure}. 
The reconstructed tertiary structure (cyan) achieves high structural fidelity relative to the original data (blue), highlighting the accuracy of our approach. 
For comparison, a structure (gray) was generated by fixing all $sp^3$ bond angles at their ideal tetrahedral value ($109^\circ28'$). 
This comparison reveals a significant structural mismatch, including the loss of several $\alpha$-helical segments. 
Such discrepancies underscore the substantial impact of even minor bond-angle variations on the global protein fold and emphasize the necessity of incorporating these degrees of freedom into our representation. 
The secondary structure representation also exhibits excellent agreement with the original data, with only negligible deviations at the termini of specific side-chain atoms. 
Quantitatively, analysis of $C_\alpha$ atom positions yields a maximum deviation of 0.26~\AA\ and an average deviation of 0.04~\AA. 
For side-chain atoms, the maximum and mean differences are 0.6~\AA\ and 0.26~\AA, respectively. 
The RMSDs for both $C_\alpha$ and heavy atoms of T1027 consistently approach zero. 
As illustrated in the lower panel of Fig.~\ref{fig:single-chiain-structure}, the reconstruction of 1l2y aligns almost perfectly with the reference structure, characterized by a $C_\alpha$ RMSD of 0.18~\AA\ and an all-heavy-atom RMSD of 0.05~\AA. 
Our analysis suggests that the remaining minor discrepancies primarily originate from slight bond-length deviations within the terminal amino groups.

\begin{figure}
\centering
\begin{subfigure}{0.45\textwidth}
\includegraphics{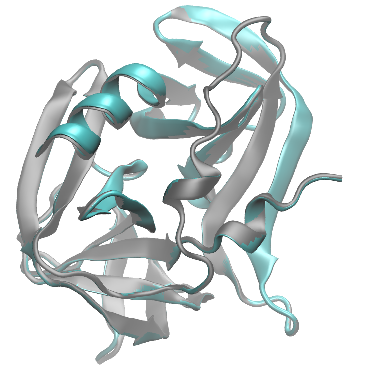}
\end{subfigure}
\begin{subfigure}{0.45\textwidth}  
\includegraphics{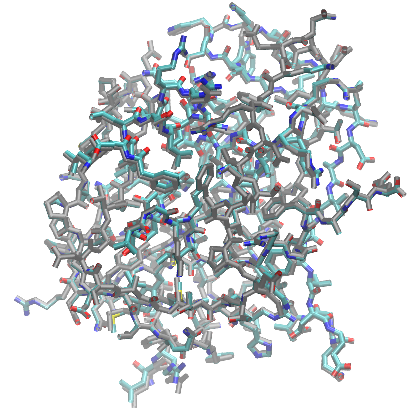}
\end{subfigure}
\begin{subfigure}{0.45\textwidth}
\includegraphics{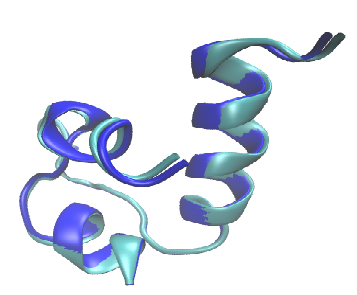}
\end{subfigure}
\caption{Multi-chain protein coordinate reconstruction. (a, c) Tertiary structure comparison and (b) side-chain consistency for 187-residue protein 3sj9 (large) and 46-residue protein 1bom (small). The native structure (blue) is compared against a reconstruction (cyan) generated using all ground-truth dihedral and bond angles.
}
\label{fig:two-chain-structure}
\end{figure}
To assess the protocol’s efficacy for multi-chain systems, we performed a reconstruction of the 187-residue, two-chain protein 3sj9~\cite{Lu2011}. 
As illustrated in Fig.~\ref{fig:two-chain-structure}, the reconstructed model achieves high structural fidelity, with the backbone and side-chain geometries closely mirroring the native configuration. 
Quantitatively, the RMSD between the reconstructed and original structures is 0.26~\AA~for backbone atoms and 0.32~\AA~for all heavy atoms.
These values signify near-native reconstruction accuracy.
Comparable performance was observed for the smaller protein 1bom, which yielded an RMSD of 0.58~\AA\ for $C_\alpha$ atoms and 0.62~\AA\ for all heavy atoms (Fig.~\ref{fig:two-chain-structure}, lower panel). 
The marginal increase in variance observed in multi-chain systems, relative to single-chain reconstructions, is primarily attributed to the initialization of the global reference frame for each individual chain. 
Because these frames are established based on the first few atoms of an existing template—the positions of which may vary slightly from the dynamic, true coordinates—minor structural shifts can occur. 
However, the numerical results confirm that these deviations are statistically insignificant.
\subsection{\label{sec:application-structure}Trajectory generation}
\subsubsection{Universal Transformer Application}
To evaluate the code, we used two proteins as templates: a 20-amino-acid single-chain protein, 1l2y~\cite{Neidigh2002}, and a 46-amino-acid two-chain protein 1bom~\cite{Nagataetal1995}.
Molecular dynamics simulations were carried out for 100 ns for both proteins, and trajectory frames were saved at 0.1 ns intervals.
The trajectories of 1l2y and 1bom are put together for training one unified model. Each trajectory consisted of 1000 structures.
\par
Based on the analysis presented in the Methods section, the minimum tensor dimension required for representing each amino acid is 40.
However, to ensure compatibility with different amino acid configurations, we use a tensor dimension of 300.
An adaptive learning rate scheme is implemented, adjusting the learning rate based on the training loss.
Our model employs two Transformer layers.
The dense layers within 
$\mathbb{D}_S$ and $\mathbb{D}_E$ each contain a single hidden layer with length 3200.
Following our previous work, the LeakyReLu activation function is consistently used in all Transformer layers and the dense layers of $\mathbb{D}_S$ and $\mathbb{D}_E$.
\par
After training the model, we generate trajectories by iteratively applying the learned propagator, $\mathbb{F}_0$.
We apply a structure near the starting of the training MD trajectory as the initial structure.
Each subsequent frame is then generated conditioned only on the previous frame's coordinates.
This generative process achieves a speedup of approximately 10,000 times relative to standard molecular dynamics computations.
To assess the accuracy of the generated trajectory, we compare it to the original MD trajectory using the RMSD of the $C_\alpha$ atoms.
The RMSD of $C_\alpha$ atoms provides a sensitive measure of structural similarity, as it is highly responsive to even small changes in individual amino acid conformation.
It is important to emphasize that we use only one frame of the MD data to initiate the generation process; all subsequent frames are de novo generated.
This constitutes a stringent test, as errors in propagation can accumulate, and any single incorrect step can lead to significant deviations in the generated trajectory.
\par
\begin{figure}
\begin{subfigure}{0.45\textwidth}  
\includegraphics[width=\textwidth]{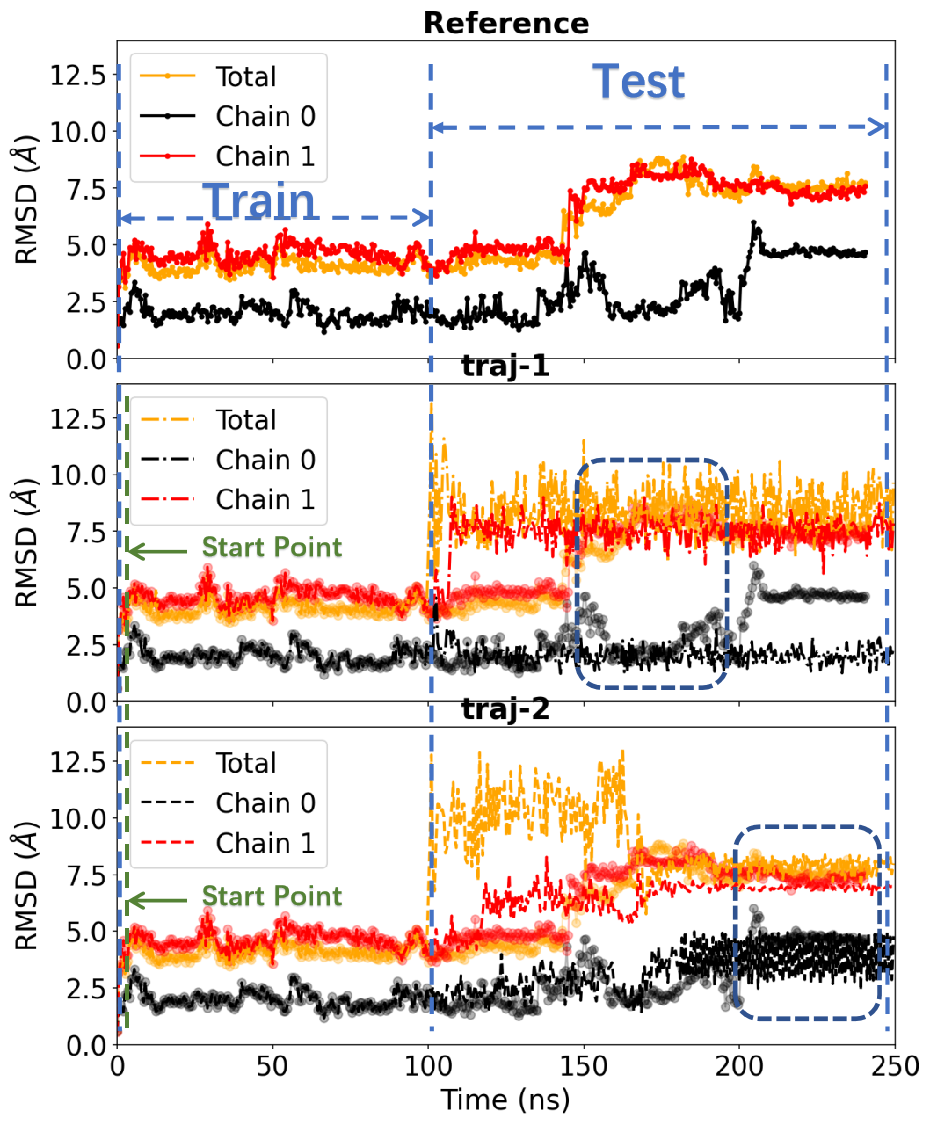}
\end{subfigure}
\begin{subfigure}{0.45\textwidth}  
\includegraphics[width=\textwidth]{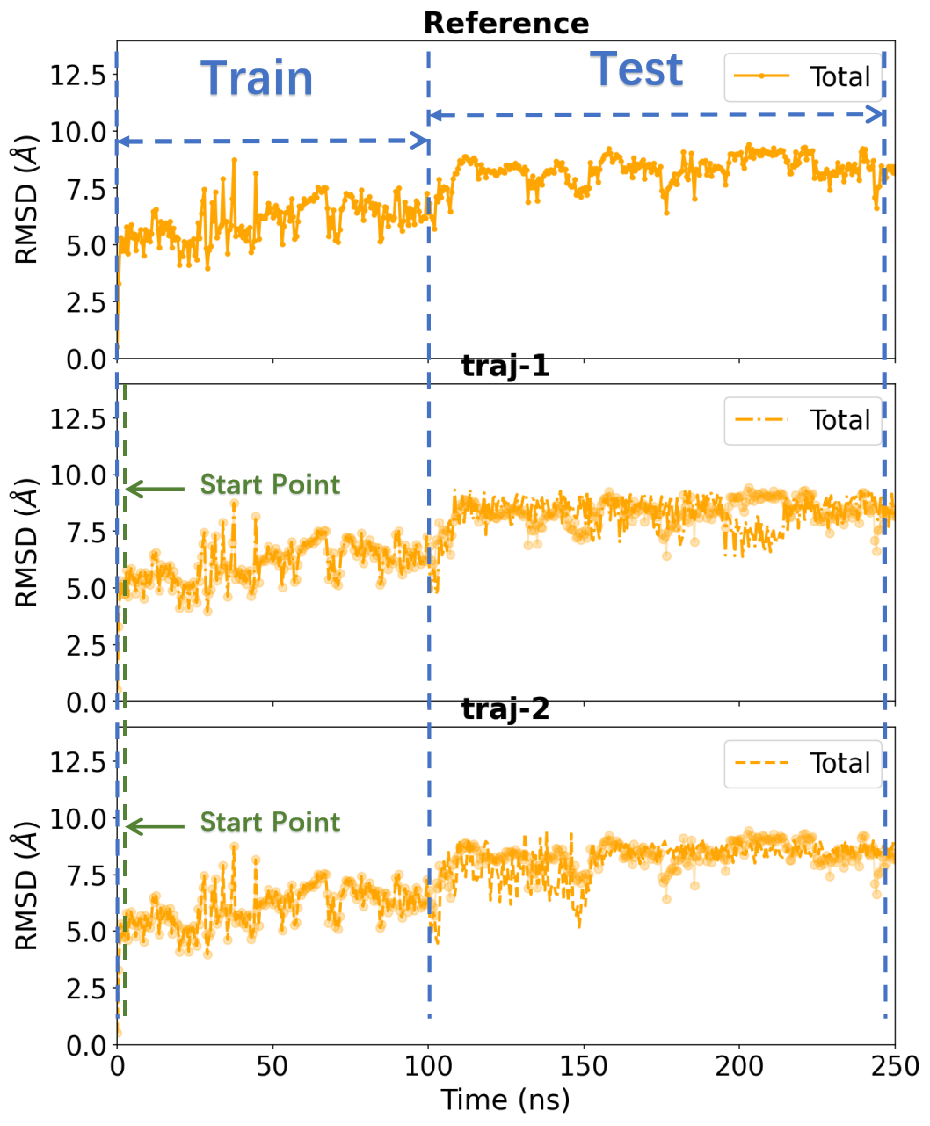}
\end{subfigure}
\caption{Comparison of MD and Transformer-generated RMSD profiles for 1BOM and 1L2Y. RMSD is plotted for the total protein (orange), chain 1 (black), and chain 2 (red). Top panels show original MD trajectories; lower panels display generated trajectories (0-250 ns, dotted lines) superimposed on semi-transparent MD data (solid lines). The generator was trained on the initial 100 ns of MD data and initialized using a structure near the training starting point.
}
\label{fig:1bom-1l2y-rmsd}
\end{figure}

The RMSD profiles for 1bom and 1l2y are presented in Fig.~\ref{fig:1bom-1l2y-rmsd}. Within the first 100~ns training window, the predicted RMSD values align closely with the reference data, demonstrating robust interpolation of the trajectory.
When extrapolating beyond the training domain, the model maintains strong agreement with the reference RMSD profiles. 
For 1bom, the individual chains exhibit consistent alignment in the test data. Specifically, in traj-1, both chains and the total protein performance show high fidelity between 150--200~ns, with this agreement persisting through the 200--250~ns interval. The consistency between individual and global performance in the test data validates the model's accuracy in generating MD trajectories and its sampling efficiency. Even within the more challenging 100--150~ns region, satisfactory agreement is maintained for at least one individual chain.
For the single-chain protein 1l2y, the predicted RMSD profile accurately tracks the MD trajectory across the 100--250~ns range. 

Notably, the RMSD patterns in the test phase differ significantly from those in the training set. For 1bom, the plateaus for chain 1 and the total protein (approximately 8~\AA) are higher than the 5~\AA\ plateau observed during training; similarly, the 5~\AA\ plateau for chain 0 exceeds the 2.5~\AA\ training value. In the case of 1l2y, the test plateau reaches approximately 8~\AA, which is higher than the training RMSD profiles. These results highlight the model's extrapolative capacity and suggest that it has successfully learned the fundamental internal forces governing protein dynamics.

\subsubsection{Comparision: Protein specific DNN+RealNVP framework}
\begin{figure}
\centering
    \includegraphics[width=0.45\textwidth]{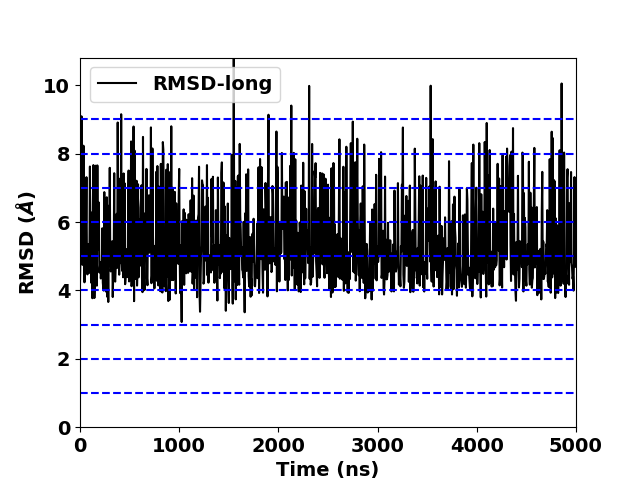}
\caption{Simulation with first frame of the testing data. Noise coefficient is and a constant temperature simulation is performed. Its lowest RMSD is around 3.45~\AA, lower than the lower limit 3.83~\AA~from the training data.}
\label{fig:long-simulation}
\end{figure}

We include a basic DNN+RealNVP framework for comparison. In this architecture, the input vector dimensionality is protein-dependent, which limits its generalizability.
We observe that while a simple DNN can easily generate short trajectories (20 ns) that closely match the original MD data (Fig.~\ref{fig:MD-results-T1027} in the appendix), the inclusion of a noise component is essential for continuous trajectory generation.
In long-duration simulations, we found that the generated trajectories consistently exhibit RMSD variations between 
3-9~\AA. This significantly exceeds the 4-6~\AA~range observed in the original test data.
While this increased variance can facilitate structural extrapolation (Fig.~\ref{fig:long-simulation}), the resulting trajectory dynamics deviate substantially from the reference MD behavior.

\subsection{\label{sec:noise-propagation}Propagation with Noise}

As demonstrated in previous sections, a low dropout rate ($10^{-6}$) introduces sufficient variance to successfully replicate MD trajectories. Here, we demonstrate that the dropout parameter can serve as a physical proxy for temperature within the MD simulation framework.

\begin{figure}
\begin{subfigure}{0.45\textwidth}  
\includegraphics[width=\textwidth]{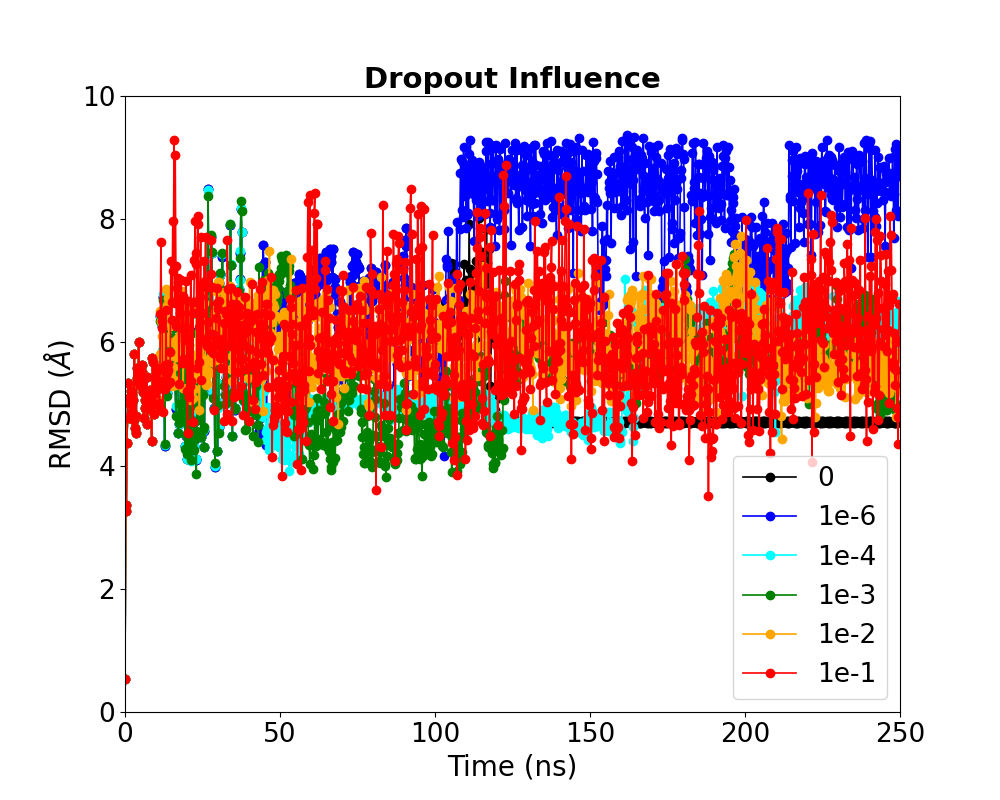}
\end{subfigure}
\begin{subfigure}{0.45\textwidth}  
\includegraphics[width=\textwidth]{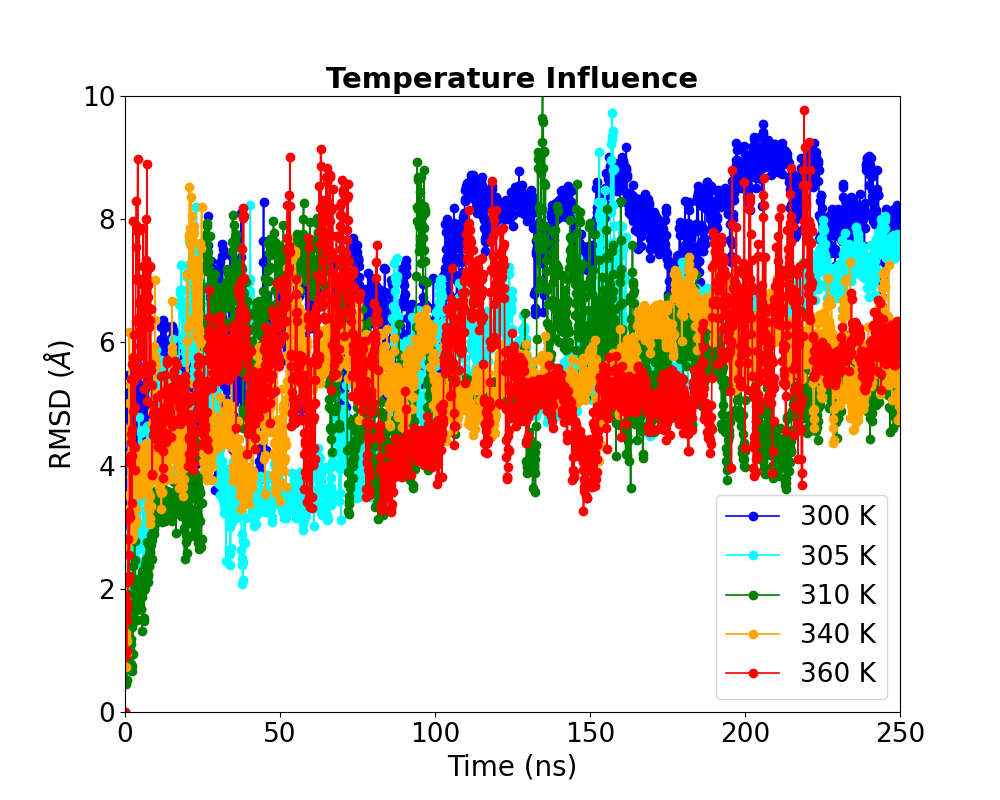}
\end{subfigure}
\begin{subfigure}{0.45\textwidth}  
\includegraphics[width=\textwidth]{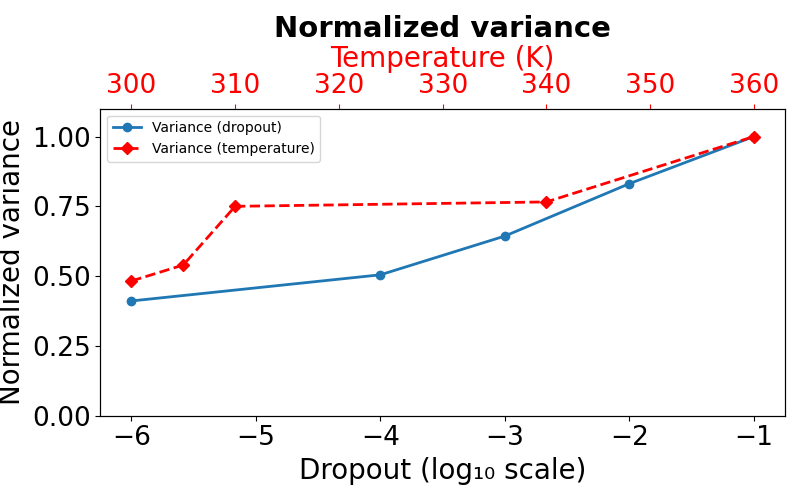}
\end{subfigure}
\caption{Comparison of model-generated and Gromacs MD trajectories for 1L2Y. (Top) 250 ns RMSD profiles for models trained with varying dropout parameters (0, 1e-5, 1e-4, 1e-3, 1e-2 and 1e-1), starting from the initial reference frame. (Middle) Standard RMSD profiles for Gromacs MD data at 300 K, 305 K, 310 K 340 K and 360 K for comparison of conformational stability.(Lower) The computed variances at of RMSD at different temperatures.
}
\label{fig:1l2y-noise-tem}
\end{figure}

As shown in the upper panel of Fig.~\ref{fig:1l2y-noise-tem}, as the dropout increases from 0 to 0.1, there is a corresponding rise in RMSD variance. With a dropout of 0 (represented by the black line), the RMSD remains nearly static between 100 and 250~ns, highlighting that stochastic deactivation is essential for facilitating realistic molecular dynamics. As the dropout rate increases incrementally, the RMSD variance expands from approximately 1~\AA~(green for 1e-5 and blue for 1e-4) to 4~\AA~(orange for 1e-3), ultimately exceeding 4~\AA~(1e-1). A similar trend is observed in all-atom MD trajectories across a temperature range of 300~K to 360~K, as depicted in the middle panel of Fig.~\ref{fig:1l2y-noise-tem}. 
Specifically, the RMSD increases from less than 1~\AA~at 300~K (blue) to over 4~\AA~at 360~K.
Calculated via Geary’s C~\cite{Geary1954}, this metric reflects the difference between a value and its immediate neighbors. As shown in the lower panel of Fig.~\ref{fig:1l2y-noise-tem}, the normalized local RMSD variance grows as both the model's dropout rate and the MD temperature increase.

Consequently, the dropout rate can be precisely calibrated to simulate MD at various target temperatures in practice. We hypothesize that with a sufficiently comprehensive and diverse MD training dataset covering the relevant configuration space, noise can be more effectively integrated to promote the exploration of novel or transition states.


\section{\label{sec:conclusion}Conclusion and outlook}
This research introduces a robust framework that leverages artificial intelligence to facilitate high-fidelity, CG simulations of protein dynamics. 
By establishing a bidirectional mapping between CVs and three-dimensional atomic coordinates through tree-like coefficients, we demonstrate a near-native reconstruction of structural topology. 
This approach was rigorously validated across diverse systems, ranging from small peptides like 1l2y and 1bom to complex proteins such as 3sj9 and T1027.

Central to this work is the novel representation of CVs as language-like sequences, which allows the protein propagation problem to be framed as a sequence-to-sequence task. 
By employing a unified Transformer-based architecture as a structural propagator, we achieved a $10^4$-fold speedup in MD simulations compared to conventional all-atom methods. 
The model demonstrates significant generalizability, successfully performing both interpolation and extrapolation on unseen test data, signaling a major step toward a truly universal protein propagator.

Looking forward, several promising directions emerge:
\begin{itemize}
    \item \textbf{Toward a Foundation Model of Protein Dynamics:} While this study validates the framework on specific systems, the Transformer-based architecture is inherently scalable. Future iterations will involve training on massive, multi-microsecond trajectory datasets to develop a foundation model capable of predicting the dynamics of any protein sequence without further training.
    \item \textbf{High-Throughput Kinetic Screening:} The $10^4$-fold speedup provides an unprecedented opportunity for drug discovery. This framework could be utilized to simulate thousands of ligand-protein binding events in the time it currently takes to simulate a single system, allowing researchers to prioritize candidates based on binding kinetics rather than just static docking scores.
    \item \textbf{Real-time Structural Refinement:} Integration with experimental techniques such as cryo-electron microscopy (cryo-EM) and nuclear magnetic resonance (NMR) could allow for real-time structural refinement. The model's ability to rapidly propagate conformations could help bridge the gap between static experimental snapshots and the underlying dynamic ensemble.
    \item \textbf{Multiscale Integration:} This work lays the foundation for bridging molecular-level dynamics with macroscopic biological phenomena. By incorporating these rapid propagators into larger multiscale frameworks, we can begin to simulate cellular environments where the structural integrity of individual proteins is maintained across vast time and length scales.
\end{itemize}

\appendix
\gdef\thefigure{\thesection.\arabic{figure}}
\section{Cartesian coordinates and collective variables}

\subsection{Expanded Coordinate Transformation Matrix}\label{sec:coord-transform}
The explicit operating matrix $\hat{R}(\hat{u},\theta)$ for rotating an angle $\theta$ along a normalized rotation vector $\vec{u}=[\vec{u}_x,\vec{u}_y,\vec{u}_z]^T$ is:
\begin{equation}
    \begin{bmatrix}
\vec{u}_x^2(1-\text{c}\theta)+\text{c}\theta & \vec{u}_x\vec{u}_y(1-\text{c}\theta)-\vec{u}_z\text{s}\theta  & \vec{u}_x\vec{u}_z(1-\text{c}\theta)+\vec{u}_y\text{s}\theta\\ 
\vec{u}_x\vec{u}_y(1-\text{c}\theta)+\vec{u}_z\text{s}\theta & \vec{u}_y^2(1-\text{c}\theta)+\text{c}\theta  & \vec{u}_y\vec{u}_z(1-\text{c}\theta)-\vec{u}_x\text{s}\theta\\ 
\vec{u}_x\vec{u}_z(1-\text{c}\theta)-\vec{u}_y\text{s}\theta & \vec{u}_y\vec{u}_z(1-\text{c}\theta)+\vec{u}_x\text{s}\theta  & 
\vec{u}_z^2(1-\text{c}\theta)+\text{c}\theta\end{bmatrix}
\end{equation}
where $\text{s}\theta=\sin\theta$ and $\text{c}\theta=\cos\theta$.

\subsection{Example amino acid structure}
\setcounter{figure}{0}
\begin{figure}
\centering
    \includegraphics[width=0.45\textwidth]{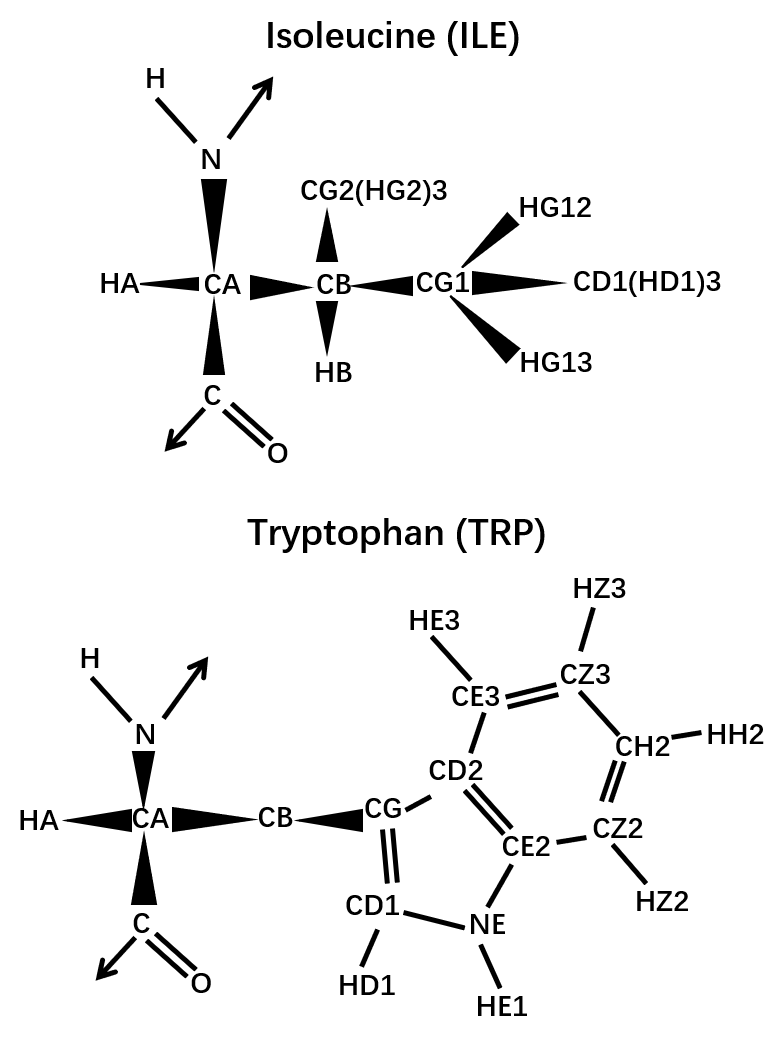}
\caption{The illustration of structures of ILE (upper panel) and TRP (lower panel).}
\label{fig:ile-trp}
\end{figure}
\subsection{Amino Acid hierarchy}
\setcounter{figure}{0}
\begin{table}[]
\caption{
Protein Structure Hierarchy (Ignoring Common N Atoms).
This figure depicts the hierarchical representation of a protein structure, with each box representing a group of atoms sharing a local coordinate frame.
The hierarchy starts with $C_\alpha$ atoms in the first column, and progressively incorporates neighboring atoms in subsequent columns (increasing depth, show as $0,1,2\cdots 6$) in the first row.}
\label{tab:amino-hierarchy}
\begin{tabular}{lccccccc}
\hline
Amino                & 0                   & 1                   & 2       & 3                                                                             & 4                                                 & 5  & 6       \\ \hline
ALA                  & CA                  & CB                  &         &                                                                               &                                                   &    &         \\ \hline
                     &                     & C                   & O       &                                                                               &                                                   &    &         \\ \hline
\multirow{2}{*}{VAL} & \multirow{2}{*}{CA} & CB                  & CG1,CG2 &                                                                               &                                                   &    &         \\ \cline{3-8} 
                     &                     & C                   & O       &                                                                               &                                                   &    &         \\ \hline
\multirow{2}{*}{LEU} & \multirow{2}{*}{CA} & CB                  & CG      & CD1,CD2                                                                       &                                                   &    &         \\ \cline{3-8} 
                     &                     & C                   & O       &                                                                               &                                                   &    &         \\ \hline
GLY                  & CA                  & C                   & O       &                                                                               &                                                   &    &         \\ \hline
\multirow{3}{*}{ILE} & \multirow{3}{*}{CA} & \multirow{2}{*}{CB} & CG1     &                                                                               &                                                   &    &         \\ \cline{4-8} 
                     &                     &                     & CG2     & CD                                                                            &                                                   &    &         \\ \cline{3-8} 
                     &                     & C                   & O       &                                                                               &                                                   &    &         \\ \hline
\multirow{2}{*}{MET} & \multirow{2}{*}{CA} & CB                  & CG      & SD                                                                            & CE                                                &    &         \\ \cline{3-8} 
                     &                     & C                   & O       &                                                                               &                                                   &    &         \\ \hline
\multirow{2}{*}{TRP} & \multirow{2}{*}{CA} & CB                        & \begin{tabular}[c]{@{}c@{}}CG,CD1,CD2\\ NE1,CE2\\ CE3,CZ2\\ CZ3,CH2\end{tabular}& &                                                   &    &         \\ \cline{3-8} 
                     &                     & C                   & O       &                                                                               &                                                   &    &         \\ \hline
\multirow{2}{*}{PHE} & \multirow{2}{*}{CA} & CB                  & CG      & \begin{tabular}[c]{@{}c@{}}CD1,CD2\\ CE1,CE2\\ CZ\end{tabular}                &                                                   &    &         \\ \cline{3-8} 
                     &                     & C                   & O       &                                                                               &                                                   &    &         \\ \hline
\multirow{2}{*}{PRO} & \multirow{2}{*}{CA} & CB                  & CG,CD   &                                                                               &                                                   &    &         \\ \cline{3-8} 
                     &                     & C                   & O       &                                                                               &                                                   &    &         \\ \hline
\multirow{2}{*}{SER} & \multirow{2}{*}{CA} & CB                  & OG      &                                                                               &                                                   &    &         \\ \cline{3-8} 
                     &                     & C                   & O       &                                                                               &                                                   &    &         \\ \hline
\multirow{2}{*}{CYS} & \multirow{2}{*}{CA} & CB                  & SG      &                                                                               &                                                   &    &         \\ \cline{3-8} 
                     &                     & C                   & O       &                                                                               &                                                   &    &         \\ \hline
\multirow{2}{*}{ASN} & \multirow{2}{*}{CA} & CB                  & CG      & OD1,ND2                                                                       &                                                   &    &         \\ \cline{3-8} 
                     &                     & C                   & O       &                                                                               &                                                   &    &         \\ \hline
\multirow{2}{*}{GLN} & \multirow{2}{*}{CA} & CB                  & CG      & CD                                                                            & OE1,OE2                                           &    &         \\ \cline{3-8} 
                     &                     & C                   & O       &                                                                               &                                                   &    &         \\ \hline
\multirow{2}{*}{THR} & \multirow{2}{*}{CA} & CB                  & OG1,CG2 &                                                                               &                                                   &    &         \\ \cline{3-8} 
                     &                     & C                   & O       &                                                                               &                                                   &    &         \\ \hline
\multirow{2}{*}{TYR} & \multirow{2}{*}{CA} & CB                  & CG      & \begin{tabular}[c]{@{}c@{}}CD1,CE1\\ CZ,OH\\ CE2,CD2\end{tabular}             &                                                   &    &         \\ \cline{3-8} 
                     &                     & C                   & O       &                                                                               &                                                   &    &         \\ \hline
\multirow{2}{*}{ASP} & \multirow{2}{*}{CA} & CB                  & CG      & \begin{tabular}[c]{@{}c@{}}OD1\\ OD2\end{tabular}                             &                                                   &    &         \\ \cline{3-8} 
                     &                     & C                   & O       &                                                                               &                                                   &    &         \\ \hline
\multirow{2}{*}{GLU} & \multirow{2}{*}{CA} & CB                  & CG      & CD                                                                            & \begin{tabular}[c]{@{}c@{}}OE1\\ OE2\end{tabular} &    &         \\ \cline{3-8} 
                     &                     & C                   & O       &                                                                               &                                                   &    &         \\ \hline
\multirow{2}{*}{LYS} & \multirow{2}{*}{CA} & CB                  & CG      & CD                                                                            & CE                                                & NZ &         \\ \cline{3-8} 
                     &                     & C                   & O       &                                                                               &                                                   &    &         \\ \hline
\multirow{2}{*}{ARG} & \multirow{2}{*}{CA} & CB                  & CG      & CD                                                                            & NE                                                & CZ & NH1,NH2 \\ \cline{3-8} 
                     &                     & C                   & O       &                                                                               &                                                   &    &         \\ \hline
\multirow{2}{*}{HIS} & \multirow{2}{*}{CA} & CB                  & CG      & \begin{tabular}[c]{@{}c@{}}ND1,CD2\\ CE1,NE2\end{tabular}                     &                                                   &    &         \\ \cline{3-8} 
                     &                     & C                   & O       &                                                                               &                                                   &    &         \\ \hline
                     &                     &                     &         &                                                                               &                                                   &    &        
\end{tabular}
\end{table}

\subsection{Frame Normalization Matrices}\label{sec:frame-normalization}
For consistency across chain blocks in the Transformer model, the relative translation origins and rotational angles of each chain are transformed into the sine-cosine format. The explicit transformations are:
\begin{equation}\label{eq:frame-prop-origin}
\begin{split}
&\vec{T}^T=\\
&\begin{bmatrix}
\frac{O^1_x}{O_\text{max}} & \cdots &\frac{O^C_z}{O_\text{max}} & \vec{0}_{L-3C} & \sqrt{1-(\frac{O^1_x}{O_\text{max}}})^2 &\cdots &\sqrt{1-(\frac{O^C_z}{O_\text{max}}})^2 & \vec{0}_{L-3C} \\
\end{bmatrix} 
\end{split}
\end{equation}
and
\begin{equation}\label{eq:frame-prop-angle} 
\vec{\theta}_f^T=\begin{bmatrix}
    \cos{\varphi^1_1} & \cdots & \cos{\varphi^C_3} & \vec{0}_{L-3C} & \sin{\varphi^1_1} & \cdots & \sin{\varphi^C_3} & \vec{0}_{L-3C} 
\end{bmatrix}
\end{equation}
where $O_{\text{max}}$ is a value exceeding any possible coordinate within the simulation box.

\section{SDE with CVs}\label{sec:SDE-CV-app}
In this section, we derive the relevant formulas introduced in Sec.~\ref{sec:SDE-CV}.
To simplify notation, we consistently employ $\mathbb{F}$ to represent the drift force function regardless of input variables.
\subsection{SDE overview}\label{sec:SDE-overview}
The noise term in Eq.~\ref{eq:general-SDE} can typically be expressed as a single Gaussian noise term.
The equation can be reformulated as follows:
\begin{equation}\label{eq:general-SDE-transform}
\begin{split}
    \frac{dx(\tau)}{d\tau}&=f(x(\tau))+\sum_\alpha g_\alpha(x(\tau))\xi_\alpha\\
    &= f(x(\tau))+\xi(0,\sigma^2(x(\tau)))\\
    \sigma(x(\tau))&\equiv \sqrt{\sum_\alpha g_\alpha^2 (x(\tau))}
\end{split},
\end{equation}
where $\xi(\mu,\sigma^2(x(\tau)))$ is a random variable drawn from a Gaussian distribution with mean $\mu$ and variance $\sigma^2(x(\tau))$.
Discretizing time with a constant interval of $\Delta \tau$, the position at the subsequent time step can be expressed as 
\begin{equation}
\begin{split}
    x(\tau+\Delta \tau)
    &=x(\tau)+f(x(\tau))\Delta \tau+\Delta \tau\xi(0,\sigma^2(x(\tau)))\\
    &=F_0(x(\tau))+\xi(0,(\sigma(x(\tau))\Delta \tau)^2)\\
    F_0(x)&\equiv x + f(x)\Delta \tau
\end{split}.
\end{equation}
Here, $\mathbb{F}_0$ denotes the true drift force component,
 while $\xi(0,\sigma^2(x(\tau))$ represents the Gaussian noise.
It can be written as
\begin{equation}\label{eq:angle-sde-simple}
    \vec{\theta}_{t+1}=\mathbb{F}_0(\vec{\theta}_t)+\xi(0,\sigma(\vec{\theta}_{t})).
\end{equation}
for our representations using angles $\theta$.
\par
Similarly, for the subsequent two steps can be expressed as
\begin{equation}
\begin{split}
    \vec{\theta}_{t+2}&=\mathbb{F}_0(\vec{\theta}_{t+1})+\xi(0,\sigma^2_{t+1})\\
    &=
\mathbb{F}_0\left(\mathbb{F}_0(\vec{\theta}_{t})+\xi(0,\sigma^2_{t})\right)+\xi(0,\sigma^2_{t+1})\\
&\approx 
\mathbb{F}_0\circ\mathbb{F}_0(\vec{\theta}_{t})+\mathbb{F}_0^{\prime}(\mathbb{F}_0(\vec{\theta}_{t}))\xi(0,\sigma^2_t)+\xi\left(0,\sigma^2_{t+1}\right)\\
&=\mathbb{F}_0\circ\mathbb{F}_0(\vec{\theta}_{t})+\xi(0,\sigma^2_{t,2})\\
&\sigma_{t,2}\equiv\sqrt{\left[\mathbb{F}_0^{\prime}(\mathbb{F}_0(\vec{\theta}_{t}))\right]^2\sigma^2_{t} + \sigma^2_{t+1}}
\end{split}.
\end{equation}
Further derivation demonstrates that the coordinates at any arbitrary time step, denoted as $t+i$, can be expressed as the drift force originating from $t$ and a zero-mean Gaussian noise:
\begin{equation}\label{eq:sde-drif-force-angle}
\begin{split}
\vec{\theta}_{t+i}&=
\overset{i}{\overbrace{\mathbb{F}_0\circ\mathbb{F}_0\cdots\circ\mathbb{F}_0}}(\vec{\theta}_{t})+\xi\left(0,\sigma^2_{t,i}\right)\\
&=\mathbb{F}_0^i(\vec{\theta}_t)+\xi\left(0,\sigma^2_{t,i}\right)\\
\mathbb{F}_0^i&\equiv \overset{i}{\overbrace{\mathbb{F}_0\circ\mathbb{F}_0\cdots\circ\mathbb{F}_0}}
\end{split}.
\end{equation}
Using $S_t$ for notional brevity, Eq.~\ref{eq:sde-drif-force-angle} becomes Eq.~\ref{eq:sde-network-formula}.
\subsection{Drift force}\label{sec:SDE-drift}
We will show Eq.~\ref{eq:loss-general} helps learns the true drift force of the SDE.
A subset of loss function in Eq.~\ref{eq:loss-general} for predicting the subsequent coordinate $S_{t+i}$ based on current time stamp $S_t$ for any time indexes $T_t=\left\{t_j|S_{t_j}=S_t,1\leq j\leq J\right\}$ can be written as
\begin{equation}\label{eq:loss-general-for-drift}
    \begin{split}
L^i_{T_t}&=\frac{1}{J}\sum_{j=1}^{J}\left \Vert S_{t+i}^j-S_{t+i,0}^j\right\Vert^2\\
        &=\left\Vert\mathbb{F}_0^i(S_t)-\mathbb{F}^i(S_t)\right\Vert^2+\left \Vert \sigma_{t,i}^2\right \Vert\\
    \mathbb{F}_0^i&\equiv \overset{i}{\overbrace{\mathbb{F}_0\circ\mathbb{F}_0\cdots\circ\mathbb{F}_0}},\mathbb{F}^i\equiv \overset{i}{\overbrace{\mathbb{F}\circ\mathbb{F}\cdots\circ\mathbb{F}}}
    \end{split},
\end{equation}
where $j$ indexes the training data, $i$ indicates the number of steps after the current time stamp and subscript $0$ refers to the true CVs.
Initially, we will establish the equivalence of the MSE of $S$ and $\vec{\theta}$ within the loss function.
For two given $S$ values which are very close, $S_\alpha=[\cos\theta_\alpha, \sin\theta_\alpha]$ and $S_\beta=[\cos\theta_\beta, \sin\theta_\beta]$
\begin{equation}
\begin{split}
    &\left \Vert S_\alpha - S_\beta \right \Vert^2\\
    =&(\cos\theta_\beta-\cos\theta_\alpha)^2+(\sin\theta_\beta-\sin\theta_\alpha)^2\\
    =&1-2\cos(\theta_\alpha-\theta_\beta)
    =4\sin^2(\frac{\theta_\alpha-\theta_\beta}{2})\\
    \sim& (\theta_\alpha-\theta_\beta)^2.
\end{split}
\end{equation}
The MSE, used for predicting the subsequent coordinate $S_{t+i}$ based on the current time stamp $t$ for any time indexes $T_t=\left\{t_j|S_{t_j}=S_t,1\leq j\leq J\right\}$, can be expressed as
\begin{equation}\label{eq:loss-part}
    \begin{split}
L^i_{T_t}&=\frac{1}{J}\sum_{j=1}^{J}\left \Vert S^j_{t+i}-S^{j}_{t+i,0}\right\Vert^2\sim \frac{1}{J}\sum_{j=1}^{J}\left \Vert \vec{\theta}^j_{t+1} -\vec{\theta}^j_{t+1,0}\right\Vert\\
        &=\frac{1}{J}\sum_{j=1}^{J}\left\Vert\mathbb{F}_0^i(\vec{\theta}_{t})+\xi_j\left(0,\sigma_{t,i}^2\right)-\mathbb{F}^i(\vec{\theta}_{t})\right\Vert^2\\
        &=\left\Vert\mathbb{F}_0^i(\vec{\theta}_t)-\mathbb{F}^i(\vec{\theta}_t)\right\Vert^2+\frac{1}{J}\sum \left\Vert\xi_j\left(0,\sigma_{t,i}^2\right)\right\Vert^2\\
        &=\left\Vert\mathbb{F}_0^i(\vec{\theta}_t)-\mathbb{F}^i(\vec{\theta}_t)\right\Vert^2+\frac{1}{J}\sum \left\Vert\sigma_{t,i}\xi(0,1)\right\Vert^2\\
        &\approx\left\Vert\mathbb{F}_0^i(\vec{\theta}_t)-\mathbb{F}^i(\vec{\theta}_t)\right\Vert^2+\left \Vert \sigma_{t,i} \right \Vert^2\\
        &\approx \left\Vert\mathbb{F}_0^i(S_t)-\mathbb{F}^i(S_t)\right\Vert^2+\left \Vert \sigma_{t,i} \right \Vert^2\\
    \mathbb{F}_0^i&\equiv \overset{i}{\overbrace{\mathbb{F}_0\circ\mathbb{F}_0\cdots\circ\mathbb{F}_0}},\mathbb{F}^i\equiv \overset{i}{\overbrace{\mathbb{F}\circ\mathbb{F}\cdots\circ\mathbb{F}}}
    \end{split},
\end{equation}
where symbols $j$ and $i$ represent the index of the training data and the steps following the current timestamp, respectively. 
\begin{figure}
\begin{subfigure}{0.5\textwidth}
\includegraphics[width=\textwidth]{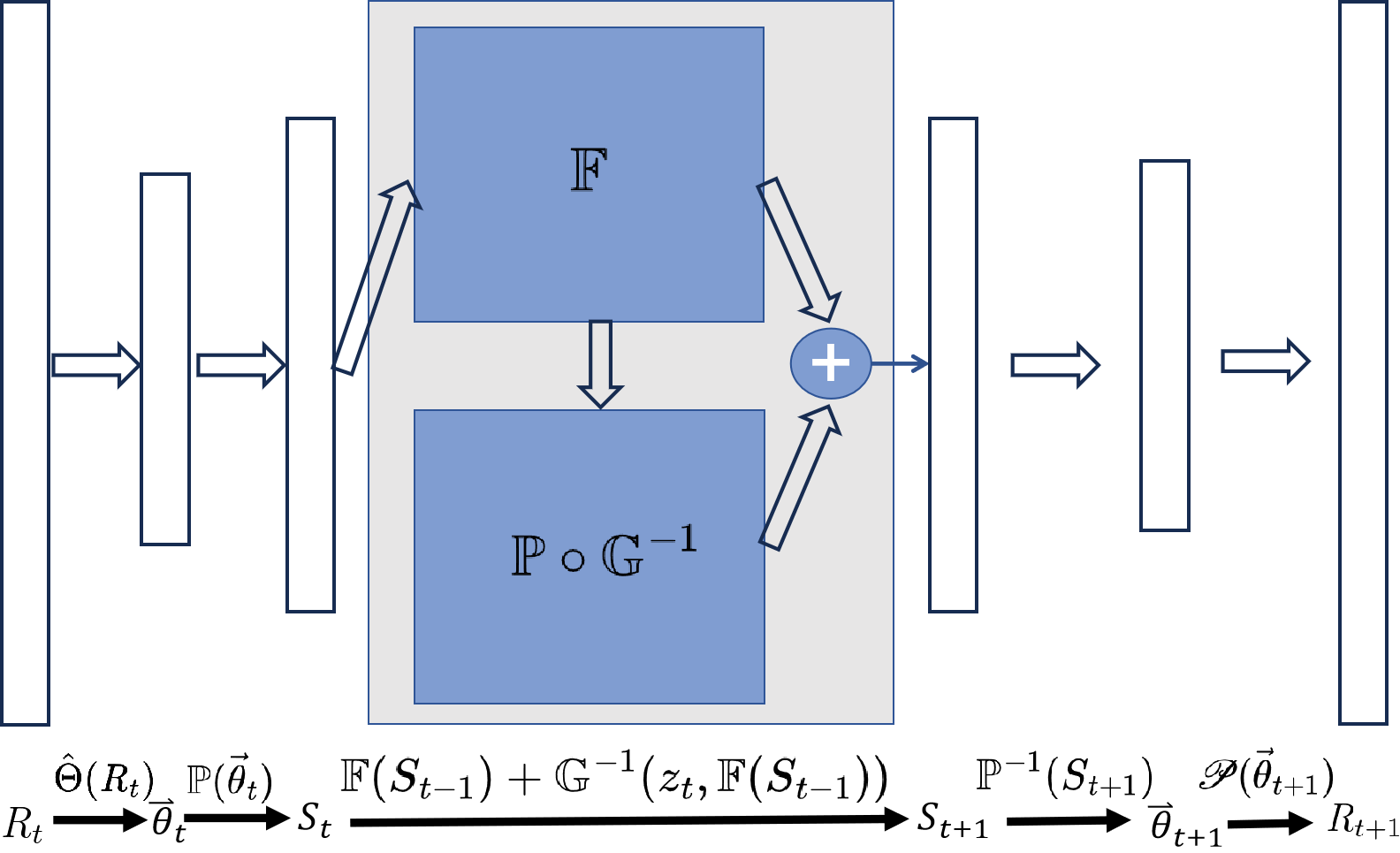}
\end{subfigure}
\begin{subfigure}{0.5\textwidth}
\includegraphics[width=\textwidth]{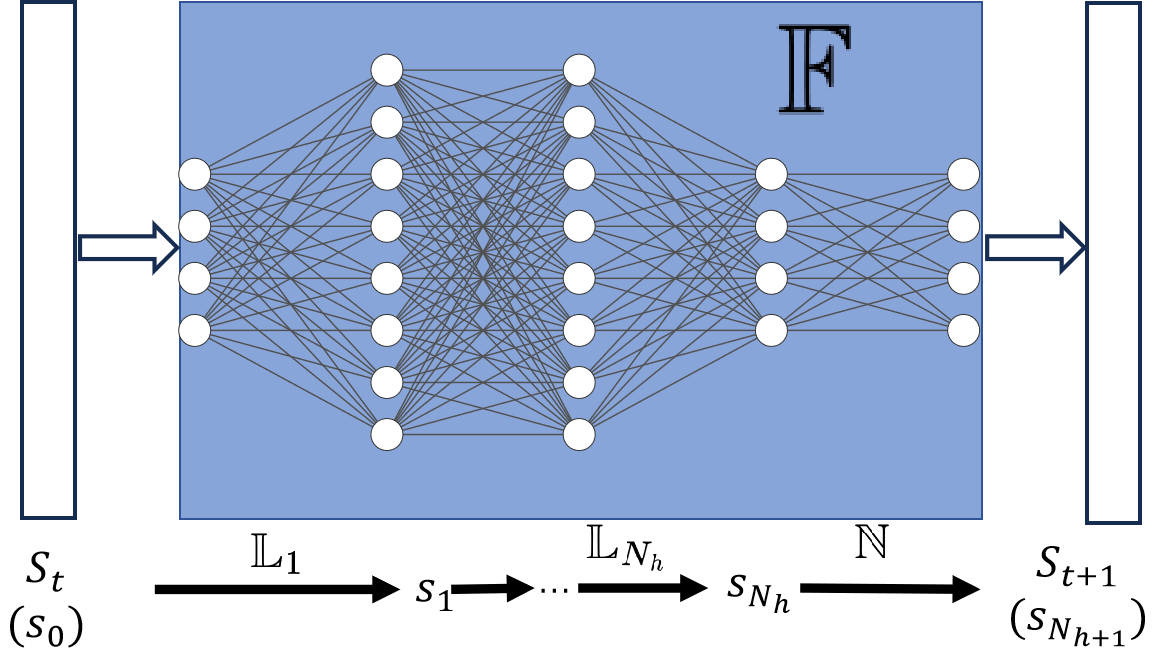}
\end{subfigure}
\begin{subfigure}{0.5\textwidth}
\includegraphics[width=\textwidth]{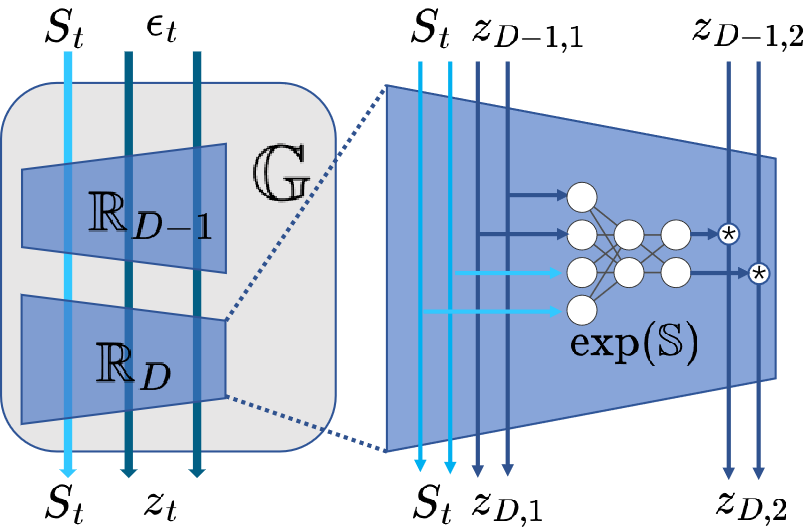}
\end{subfigure}
\caption{
Deep neural network architecture for predicting the next protein coordinate.
The upper panel illustrates the overall network structure, including modules for solving the SDE and coordinate transformation.
The middle panel details the propagator network $\mathbb{F}$, which models the drift force component.
The lower panel shows the RealNVP-based noise generator network $\mathbb{G}$, responsible for generating the noise term (see Sec.~\ref{sec:realnvp-noise} for details).}
\label{fig:network}
\end{figure}

\subsection{RealNVP-Based Noise Generator}\label{sec:realnvp-noise}

RealNVP~\cite{Dinh2017} is a versatile method for estimating probability distributions, previously employed in the Boltzmann generator~\cite{Noe2019} for identifying intermediate states in MD simulations, and in Timewarp~\cite{Klein2023} for noise term prediction in coordinate and velocity SDEs.
Here we describe the explicit noise generator used in our earlier single-chain framework~\cite{Zhu2024}.
\par
Unlike the drift force calculation, we employ angles as CVs for noise estimation, since the periodicity of angles does not affect distribution computations.
The noise term $\epsilon_t = \sigma_t\xi$ for angular noise is derived from two consecutive collective coordinates as
\begin{equation}
    \epsilon_t= \mathbb{P}^{-1}S_t-\mathbb{P}^{-1}\mathbb{F}(S_{t-1}),
\end{equation}
where $\mathbb{F}$ is the previously determined drift force.

\par
Following the RealNVP framework, a transformation $\mathbb{G}$ maps the noise term $\epsilon_t$ to a transformed variable $z_t$ following a standard normal distribution:
\begin{equation}
    \mathbb{G}(\epsilon_t,S_t)=z_t, \quad z_t\sim \mathit{N}(0,I).
\end{equation}
The network is composed of $N_h$ RealNVP layers $\mathbb{R}_D$:
\begin{equation}
\begin{split}
    \mathbb{G}=&\mathbb{R}_{N_h}\circ\cdots\circ \mathbb{R}_{1}\\
    z_D=&\mathbb{R}_D(z_{D-1},S_t),\quad z_0=\epsilon_t,\quad z_{N_h}=z_t.
\end{split}
\end{equation}
Each layer $\mathbb{R}_D$ maps:
\begin{equation}
    \mathbb{R}_D:\begin{bmatrix} z_{D,1} \\ z_{D,2} \end{bmatrix}
    = \begin{bmatrix} z_{D-1,1} \\ z_{D-1,2}\odot \exp \left(\mathbb{S}_D(z_{D-1,1},S_t)\right) \end{bmatrix},
\end{equation}
with inverse
\begin{equation}
    \mathbb{R}_D^{-1}:\begin{bmatrix} z_{D-1,1} \\ z_{D-1,2} \end{bmatrix}
    = \begin{bmatrix} z_{D,1} \\ z_{D,2}\odot \exp \left(-\mathbb{S}_D(z_{D,1},S_t)\right) \end{bmatrix},
\end{equation}
and scale network
\begin{equation}
    \mathbb{S}_D(z_{D-1,1},S_t)= \mathbb{L}_{N_D}\circ\cdots\circ \mathbb{L}_{1}
    \begin{bmatrix} \mathcal{P}^{-1}(S_t) \\ z_{D-1,1} \end{bmatrix},
\end{equation}
where $\mathbb{L}_d$ shares the form of Eq.~\ref{eq:linear-transform}.
Unlike standard RealNVP, we include only a scale term $\mathbb{S}$ (omitting the translation term), justified by the zero-mean property of $\epsilon_t$ and the incorporation of $S_t$ in the scale network.
The positions of $z_{D,1}$ and $z_{D,2}$ are swapped after each layer to enhance model capacity.

\par
The probability of the generated noise is
\begin{equation}
    p(\epsilon_t,S_t)=p(z_t)\exp\left(\sum_{D=1}^{N_h}\mathbb{S}_D(z_{D,1},S_t)\right).
\end{equation}
Network parameters are optimized by maximizing the acceptance ratio
\begin{equation}
\begin{split}
    r(S_{t-1}, \tilde{S}_t)=
    \frac{\mu(\tilde{S}_t)p(S'_t,\tilde{S}_t)}{
    \mu(S'_t)p(\tilde{S}_t,S'_t)}
    =
    \frac{\mu(S'_t+\epsilon_t)p(-\epsilon_t,S'_t+\epsilon_t)}{
    \mu(S'_t)p(\epsilon_t,S'_t)},
\end{split}
\end{equation}
where $\tilde{S}_{t}=\mathbb{F}(S_{t-1})+\mathbb{P}\circ\mathbb{G}^{-1}z_t$, $S'_t=\mathbb{F}(S_{t-1})$, and $\mu(S_t)$ is the probability of a given CV configuration, learned using a standard RealNVP network~\cite{Dinh2017}.
The loss function is
\begin{equation}
\begin{split}
    L_{N}=&\frac{1}{T}\sum_{t=1}^{T}\sum_{z_t} \log r(S_{t-1}, \tilde{S}_t)\\
         =&\frac{1}{T}\sum_{t=1}^{T}\sum_{z_t}\log\frac{\mu(S'_t+\mathbb{G}^{-1}z_t)\,p(-\mathbb{G}^{-1}z_t,\,S'_t+\epsilon_t)}{
    \mu(S'_t)\,p(\mathbb{G}^{-1}z_t,\,S'_t)}.
\end{split}
\end{equation}
Training involves simultaneous sampling of random noise $z_t$ and minimization of $L_N$.
\section{Alternative training}
\subsection{Drift force only (single chain)}
\begin{figure}
\centering
\begin{subfigure}{0.5\textwidth}
    \includegraphics[width=\textwidth]{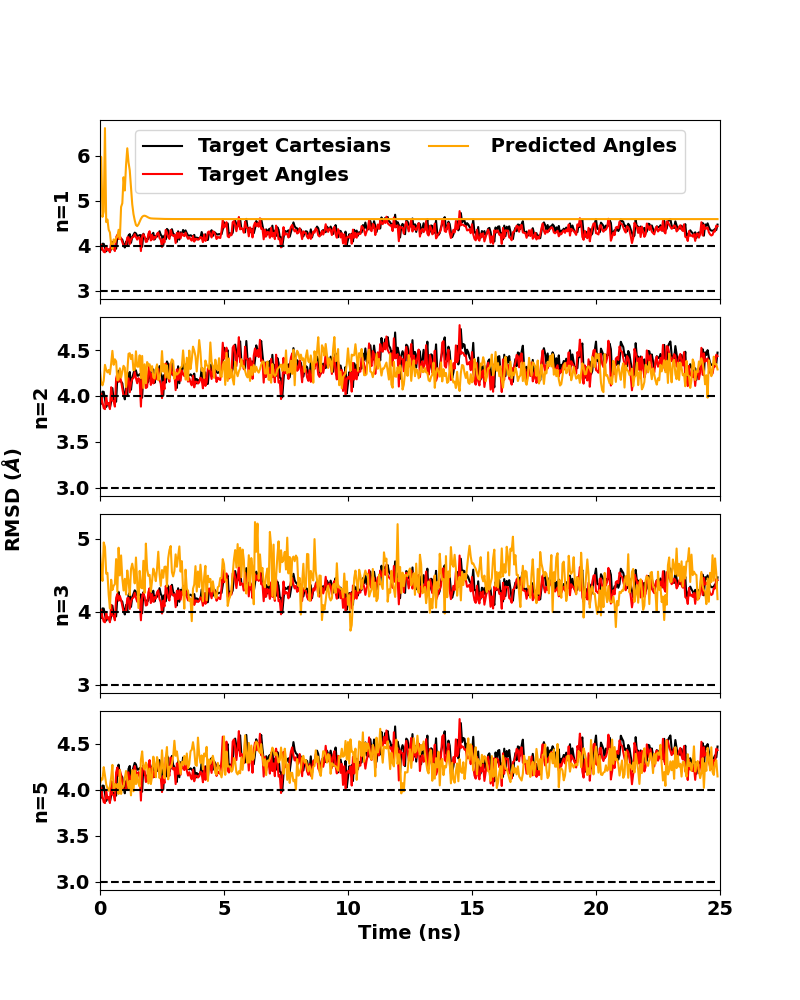}
    \caption{RMSDs for n=1, 2, 3 and 5.}
\end{subfigure}
\begin{subfigure}{0.45\textwidth}
    \includegraphics[width=\textwidth]{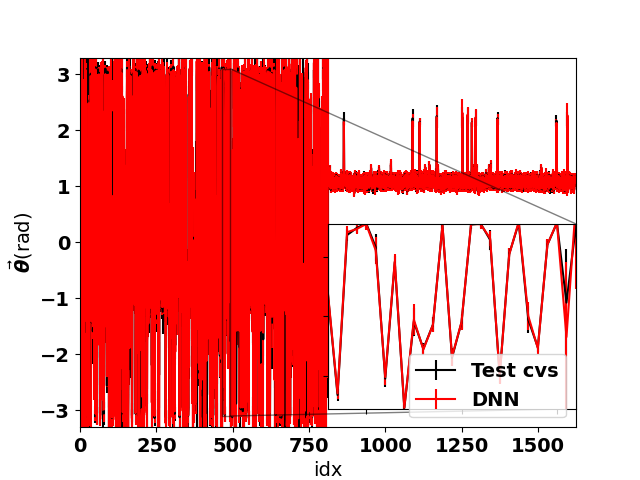}
    \caption{$\vec{\theta}$ statistics with n=2.}
\end{subfigure}
\caption{RMSD of $C_\alpha$ atoms (upper) and statistical performance of angles (lower) in the test datasets. 
Readers are redirected to CASP14 website for RMSD calculation methods. In the lower figure, the values in the nodes are the mean values and the vertical lines in the node show the standard deviations.
In the upper figure, the black line represents rmsds calculated from original trajectories, the red line represents the rmsds obtained from structures re-constructed from CVs and the orange line represents the rmsds from predicted structures.
}
\label{fig:MD-results-T1027}
\end{figure}

\bibliography{aipsamp}

@article{Zhu2021,
author = {Zhu, Jinzhen},
doi = {10.1103/PhysRevA.103.013113},
issn = {24699934},
journal = {Physical Review A},
month = {jan},
number = {1},
pages = {013113},
publisher = {American Physical Society},
title = {{Theoretical investigation of the Freeman resonance in the dissociative ionization of H2+}},
url = {https://journals.aps.org/pra/abstract/10.1103/PhysRevA.103.013113},
volume = {103},
year = {2021}
}

@article{Zhu2020,
author = {Zhu, Jinzhen and Scrinzi, Armin},
doi = {10.1103/PhysRevA.101.063407},
issn = {2469-9926},
journal = {Physical Review A},
month = {jun},
number = {6},
pages = {063407},
publisher = {American Physical Society},
title = {{Electron double-emission spectra for helium atoms in intense 400-nm laser pulses}},
url = {https://link.aps.org/doi/10.1103/PhysRevA.101.063407},
volume = {101},
year = {2020}
}

@article{Zhu2020b,
author = {Zhu, Jinzhen},
doi = {10.1103/PhysRevA.102.053109},
issn = {2469-9926},
journal = {Physical Review A},
number = {5},
pages = {053109},
publisher = {American Physical Society},
title = {{Quantum simulation of dissociative ionization of H 2 + in full dimensionality with a time-dependent surface-flux method}},
url = {https://link.aps.org/doi/10.1103/PhysRevA.102.053109},
volume = {102},
year = {2020}
}

@article{Noid2013,
author = {Noid, W. G.},
doi = {10.1063/1.4818908},
issn = {00219606},
journal = {Journal of Chemical Physics},
number = {9},
pmid = {24028092},
title = {{Perspective: Coarse-grained models for biomolecular systems}},
volume = {139},
year = {2013}
}

@article{Parsons2005,
author = {Parsons, Jerod and Holmes, J Bradley and Rojas, J Maurice and Tsai, Jerry and Strauss, Charlie E M},
doi = {10.1002/jcc.20237},
journal = {Wiley InterScience},
title = {{Practical Conversion from Torsion Space to Cartesian Space for In Silico Protein Synthesis}},
volume = {0211458},
year = {2005}
}

@article{Zhang2018,
archivePrefix = {arXiv},
arxivId = {1712.03461},
author = {Zhang, Linfeng and Wang, Han and Weinan, E.},
doi = {10.1063/1.5019675},
eprint = {1712.03461},
issn = {00219606},
journal = {Journal of Chemical Physics},
number = {12},
pmid = {29604808},
title = {{Reinforced dynamics for enhanced sampling in large atomic and molecular systems}},
volume = {148},
year = {2018}
}

@article{Jumper2021,
author = {Jumper, John and Evans, Richard and Pritzel, Alexander and Green, Tim and Figurnov, Michael and Ronneberger, Olaf and Tunyasuvunakool, Kathryn and Bates, Russ and {\v{Z}}{\'{i}}dek, Augustin and Potapenko, Anna and Bridgland, Alex and Meyer, Clemens and Kohl, Simon A.A. and Ballard, Andrew J. and Cowie, Andrew and Romera-Paredes, Bernardino and Nikolov, Stanislav and Jain, Rishub and Adler, Jonas and Back, Trevor and Petersen, Stig and Reiman, David and Clancy, Ellen and Zielinski, Michal and Steinegger, Martin and Pacholska, Michalina and Berghammer, Tamas and Bodenstein, Sebastian and Silver, David and Vinyals, Oriol and Senior, Andrew W. and Kavukcuoglu, Koray and Kohli, Pushmeet and Hassabis, Demis},
doi = {10.1038/s41586-021-03819-2},
issn = {14764687},
journal = {Nature},
number = {7873},
pages = {583--589},
pmid = {34265844},
publisher = {Springer US},
title = {{Highly accurate protein structure prediction with AlphaFold}},
url = {http://dx.doi.org/10.1038/s41586-021-03819-2},
volume = {596},
year = {2021}
}

@article{Wang2022,
archivePrefix = {arXiv},
arxivId = {2104.01620},
author = {Wang, Dongdong and Wang, Yanze and Chang, Junhan and Zhang, Linfeng and Wang, Han and E, Weinan},
doi = {10.1038/s43588-021-00173-1},
eprint = {2104.01620},
issn = {26628457},
journal = {Nature Computational Science},
number = {1},
pages = {20--29},
title = {{Efficient sampling of high-dimensional free energy landscapes using adaptive reinforced dynamics}},
volume = {2},
year = {2022}
}

@article{Liwo2001,
author = {Liwo, Adam and Czaplewski, Cezary and Pillardy, Jaroslaw and Scheraga, Harold A.},
doi = {10.1063/1.1383989},
issn = {00219606},
journal = {Journal of Chemical Physics},
number = {5},
pages = {2323--2347},
title = {{Cumulant-based expressions for the multibody terms for the correlation between local and electrostatic interactions in the united-residue force field}},
volume = {115},
year = {2001}
}

@article{Bergdorf2021,
author = {Bergdorf, Michael and Robinson-Mosher, Avi and Guo, Xinyi and Law, Ka-Hei and Shaw, David E and Shaw, D E},
number = {April},
title = {{Desmond/GPU Performance as of April 2021}},
volume = {1},
year = {2021}
}

@article{Shaw2021,
author = {Shaw, David E. and Adams, Peter J. and Azaria, Asaph and Bank, Joseph A. and Batson, Brannon and Bell, Alistair and Bergdorf, Michael and Bhatt, Jhanvi and {Adam Butts}, J. and Correi, Timothy and Dirks, Robert M. and Dror, Ron O. and Eastwoo, Michael P. and Edwards, Bruce and Even, Amos and Feldmann, Peter and Fenn, Michael and Fenton, Christopher H. and Forte, Anthony and Gagliardo, Joseph and Gill, Gennette and Gorlatova, Maria and Greskamp, Brian and Grossman, J. P. and Gullingsrud, Justin and Harper, Anissa and Hasenplaugh, William and Heily, Mark and Heshmat, Benjamin Colin and Hunt, Jeremy and Ierardi, Douglas J. and Iserovich, Lev and Jackson, Bryan L. and Johnson, Nick P. and Kirk, Mollie M. and Klepeis, John L. and Kuskin, Jeffrey S. and Mackenzie, Kenneth M. and Mader, Roy J. and McGowen, Richard and McLaughlin, Adam and Moraes, Mark A. and Nasr, Mohamed H. and Nociolo, Lawrence J. and O'Donnell, Lief and Parker, Andrew and Peticolas, Jon L. and Pocina, Goran and Predescu, Cristian and Quan, Terry and Salmon, John K. and Schwink, Carl and Shim, Keun Sup and Siddique, Naseer and Spengler, Jochen and Szalay, Tamas and Tabladillo, Raymond and Tartler, Reinhard and Taube, Andrew G. and Theobald, Michael and Towles, Brian and Vick, William and Wang, Stanley C. and Wazlowski, Michael and Weingarten, Madeleine J. and Williams, John M. and Yuh, Kevin A.},
doi = {10.1145/3458817.3487397},
isbn = {9781450384421},
issn = {21674337},
journal = {International Conference for High Performance Computing, Networking, Storage and Analysis, SC},
pages = {1--11},
title = {{Anton 3: Twenty Microseconds of Molecular Dynamics Simulation before Lunch}},
year = {2021}
}

@article{VanDerSpoel2005,
author = {{Van Der Spoel}, David and Lindahl, Erik and Hess, Berk and Groenhof, Gerrit and Mark, Alan E. and Berendsen, Herman J.C.},
doi = {10.1002/jcc.20291},
issn = {01928651},
journal = {Journal of Computational Chemistry},
number = {16},
pages = {1701--1718},
pmid = {16211538},
title = {{GROMACS: Fast, flexible, and free}},
volume = {26},
year = {2005}
}

@article{Lindahl2001,
author = {Lindahl, Erik and Hess, Berk and van der Spoel, David},
doi = {10.1007/S008940100045},
issn = {16102940},
journal = {Journal of Molecular Modeling},
number = {8},
pages = {306--317},
title = {{GROMACS 3.0: A package for molecular simulation and trajectory analysis}},
volume = {7},
year = {2001}
}

@article{Berendsen1995,
author = {Berendsen, H. J.C. and van der Spoel, D. and van Drunen, R.},
doi = {10.1016/0010-4655(95)00042-E},
issn = {00104655},
journal = {Computer Physics Communications},
number = {1-3},
pages = {43--56},
title = {{GROMACS: A message-passing parallel molecular dynamics implementation}},
volume = {91},
year = {1995}
}

@article{Hollingsworth2018,
author = {Hollingsworth, Scott A. and Dror, Ron O.},
doi = {10.1016/j.neuron.2018.08.011},
issn = {10974199},
journal = {Neuron},
number = {6},
pages = {1129--1143},
pmid = {30236283},
publisher = {Elsevier Inc.},
title = {{Molecular Dynamics Simulation for All}},
url = {https://doi.org/10.1016/j.neuron.2018.08.011},
volume = {99},
year = {2018}
}

@book{Zimmerman2016,
author = {Zimmerman, M. I. and Bowman, G. R.},
booktitle = {Methods in Enzymology},
doi = {10.1016/bs.mie.2016.05.032},
edition = {1},
issn = {15577988},
pages = {213--225},
pmid = {27497168},
publisher = {Elsevier Inc.},
title = {{How to Run FAST Simulations}},
url = {http://dx.doi.org/10.1016/bs.mie.2016.05.032},
volume = {578},
year = {2016}
}

@article{Bernardi2015,
author = {Bernardi, Rafael C. and Melo, Marcelo C.R. and Schulten, Klaus},
doi = {10.1016/j.bbagen.2014.10.019},
issn = {18728006},
journal = {Biochimica et Biophysica Acta - General Subjects},
number = {5},
pages = {872--877},
pmid = {25450171},
publisher = {Elsevier B.V.},
title = {{Enhanced sampling techniques in molecular dynamics simulations of biological systems}},
url = {http://dx.doi.org/10.1016/j.bbagen.2014.10.019},
volume = {1850},
year = {2015}
}

@article{Slavik2013,
author = {Slav{\'{i}}k, Anton{\'{i}}n},
doi = {10.1016/j.jmaa.2013.01.027},
issn = {0022247X},
journal = {Journal of Mathematical Analysis and Applications},
number = {1},
pages = {261--274},
publisher = {Elsevier Ltd},
title = {{Generalized differential equations: Differentiability of solutions with respect to initial conditions and parameters}},
url = {http://dx.doi.org/10.1016/j.jmaa.2013.01.027},
volume = {402},
year = {2013}
}

@article{Klein2023,
archivePrefix = {arXiv},
arxivId = {2302.01170},
author = {Klein, Leon and Foong, Andrew Y. K. and Fjelde, Tor Erlend and Mlodozeniec, Bruno and Brockschmidt, Marc and Nowozin, Sebastian and No{\'{e}}, Frank and Tomioka, Ryota},
eprint = {2302.01170},
number = {NeurIPS},
title = {{Timewarp: Transferable Acceleration of Molecular Dynamics by Learning Time-Coarsened Dynamics}},
url = {http://arxiv.org/abs/2302.01170},
year = {2023}
}

@article{Dinh2017,
archivePrefix = {arXiv},
arxivId = {1605.08803},
author = {Dinh, Laurent and Sohl-Dickstein, Jascha and Bengio, Samy},
eprint = {1605.08803},
journal = {5th International Conference on Learning Representations, ICLR 2017 - Conference Track Proceedings},
title = {{Density estimation using real NVP}},
year = {2017}
}

@article{Noe2019,
archivePrefix = {arXiv},
arxivId = {1812.01729},
author = {No{\'{e}}, Frank and Olsson, Simon and K{\"{o}}hler, Jonas and Wu, Hao},
doi = {10.1126/science.aaw1147},
eprint = {1812.01729},
issn = {10959203},
journal = {Science},
number = {6457},
pmid = {31488660},
title = {{Boltzmann generators: Sampling equilibrium states of many-body systems with deep learning}},
volume = {365},
year = {2019}
}

@article{Vaswani2017,
archivePrefix = {arXiv},
arxivId = {1706.03762},
author = {Vaswani, Ashish and Shazeer, Noam and Parmar, Niki and Uszkoreit, Jakob and Jones, Llion and Gomez, Aidan N. and Kaiser, {\L}ukasz and Polosukhin, Illia},
eprint = {1706.03762},
issn = {10495258},
journal = {Advances in Neural Information Processing Systems},
number = {Nips},
pages = {5999--6009},
title = {{Attention is all you need}},
volume = {2017-December},
year = {2017}
}

@article{Zhu2024,
archivePrefix = {arXiv},
arxivId = {2403.17513},
author = {Zhu, Jinzhen},
eprint = {2403.17513},
title = {{A unified framework for coarse grained molecular dynamics of proteins}},
url = {https://arxiv.org/abs/2403.17513v1},
year = {2024}
}

@article{Lu2011,
author = {Lu, Guangwen and Qi, Jianxun and Chen, Zhujun and Xu, Xiang and Gao, Feng and Lin, Daizong and Qian, Wangke and Liu, Hong and Jiang, Hualiang and Yan, Jinghua and Gao, George F.},
doi = {10.1128/jvi.00787-11},
issn = {0022-538X},
journal = {Journal of Virology},
number = {19},
pages = {10319--10331},
pmid = {21795339},
title = {{Enterovirus 71 and Coxsackievirus A16 3C Proteases: Binding to Rupintrivir and Their Substrates and Anti-Hand, Foot, and Mouth Disease Virus Drug Design}},
volume = {85},
year = {2011}
}

@article{Neidigh2002,
author = {Neidigh, Jonathan W. and Fesinmeyer, R. Matthew and Andersen, Niels H.},
doi = {10.1038/nsb798},
issn = {10728368},
journal = {Nature Structural Biology},
number = {6},
pages = {425--430},
pmid = {11979279},
title = {{Designing a 20-residue protein}},
volume = {9},
year = {2002}
}

@article{Nagataetal1995,
  author  = {Nagata, K. and Hatanaka, H. and Kohda, D. and Kataoka, H. and Nagasawa, H. and Isogai, A. and Ishizaki, H. and Suzuki, A. and Inagaki, F.},
  title   = {Three-dimensional solution structure of bombyxin-II an insulin-like peptide of the silkmoth {Bombyx} mori: structural comparison with insulin and relaxin},
  journal = {Journal of Molecular Biology},
  year    = {1995},
  volume  = {253},
  number  = {5},
  pages   = {749--758},
  doi     = {10.1006/jmbi.1995.0588}
}

@article{Huangetal2021,
  author  = {Huang, Yuanpeng Janet and Zhang, Ning and Bersch, Beate and Fidelis, Krzysztof and Inouye, Masayori and Ishida, Yojiro and Kryshtafovych, Andriy and Kobayashi, Naohiro and Kuroda, Yutaka and Liu, Gaohua and LiWang, Andy and Swapna, G. V. T. and Wu, Nan and Yamazaki, Toshio and Montelione, Gaetano T.},
  title   = {Assessment of prediction methods for protein structures determined by {NMR} in {CASP14}: Impact of {AlphaFold2}},
  journal = {Proteins: Structure, Function, and Bioinformatics},
  year    = {2021},
  volume  = {89},
  number  = {12},
  pages   = {1959--1976},
  doi     = {10.1002/prot.26246}
}

@article{Kryshtafovychetal2021,
  author  = {Kryshtafovych, Andriy and Schwede, Torsten and Topf, Maya and Fidelis, Krzysztof and Moult, John},
  title   = {Critical assessment of methods of protein structure prediction ({CASP})---{Round} {XIV}},
  journal = {Proteins: Structure, Function, and Bioinformatics},
  year    = {2021},
  volume  = {89},
  number  = {12},
  pages   = {1607--1617},
  doi     = {10.1002/prot.26237}
}

@article{Geary1954,
  author  = {Geary, R. C.},
  title   = {The Contiguity Ratio and Statistical Mapping},
  journal = {The Incorporated Statistician},
  year    = {1954},
  volume  = {5},
  number  = {3},
  pages   = {115--145},
  doi     = {10.2307/2986645}
}

\end{document}